# Systematic Review on Privacy Categorization

Paola Inverardi[a], Patrizio Migliarini[b], Massimiliano Palmiero[c]

[a]*Gran Sasso Science Institute - paola.inverardi@gssi.it*
[b]*University of L'Aquila - patrizio.migliarini@univaq.it*
[c]*University of Teramo - mpalmiero@unite.it*

**Abstract**

In the modern digital world users need to make privacy and security choices that have far-reaching consequences. Researchers are increasingly studying people's decisions when facing with privacy and security trade-offs, the pressing and time consuming disincentives that influence those decisions, and methods to mitigate them. This work aims to present a systematic review of the literature on privacy categorization, which has been defined in terms of profile, profiling, segmentation, clustering and personae. Privacy categorization involves the possibility to classify users according to specific prerequisites, such as their ability to manage privacy issues, or in terms of which type of and how many personal information they decide or do not decide to disclose. Privacy categorization has been defined and used for different purposes. The systematic review focuses on three main research questions that investigate the study contexts, i.e. the motivations and research questions, that propose privacy categorisations; the methodologies and results of privacy categorisations; the evolution of privacy categorisations over time. Ultimately it tries to provide an answer whether privacy categorization as a research attempt is still meaningful and may have a future.

## 1. Introduction

Information privacy relies on the collection and use of personal data. According to Anderson (2008), «Privacy is the ability and/or right to protect your personal information and extends to the ability and/or right to prevent invasions of your personal space [...]». This definition captures both the socio-psychological perspective, which contributes to privacy-related behaviours, and the legal perspective, which raises issues related to the right of individuals to be protected from personal information violations and unwarranted publicity [56]. In this vein, Nissenbaum [35] referred to privacy as a product of 'contextual integrity', or 'socio-technical systems', in which expectations and norms regarding the disclosure of information affect information flows.

Privacy has become increasingly important in people's everyday digital lives whenever they engage in online or offline activities. With the increased use of digital technologies, especially in terms of social network services (SNSs, e.g. Facebook, Instagram, Twitter), online shopping (e.g. Amazon, eBay), video telephony and online chats (e.g. Zoom, Skype, Meet, Whatsapp, Teams) and remote work suites (e.g. Office 365, Workspace), the understanding and regulation of digital users' privacy protection has become a challenging and important area. The use these and other modern systems (e.g. mobile health, financial apps) requires access to users' personal information and hosting devices, which offers benefits but also poses significant privacy concerns. Indeed, information technology is creating new social situations that challenge our assumptions about privacy and confidentiality, inevitably leading to discomfort, risks and mistrust [49].

In addition to a huge variety of different control systems for personal information protection, these concerns have motivated the adoption of regulations and laws to protect users, such as the European Union's General Data Protection Regulation (GDPR) [38]. However, often information platforms do not offer sufficient or proper controls to users, who are required to accept 'all-or-nothing' mechanisms in order to use the services.

Users' digital privacy categorisation has emerged in the last decades as a way to link privacy attitudes to digital behaviour. Privacy categorisation involves classifying users according to specific prerequisites, such as their ability to manage privacy issues, or in terms of which type of and how much personal information they decide to disclose. Privacy categorisation has been defined and used for different purposes. Starting with Westin's seminal work on marketing research, categorisation has been employed to improve the usability of digital technologies [18, 59, 45] and users' ability to express their privacy preferences [28].

The aim of this paper is to present a systematic review on privacy categorisation. It focuses on three main research questions:

**RQ1**: Identifying the **study contexts** that propose privacy categorisations.

**RQ2**: Investigating the **methodologies** and **results** of privacy categorisations.

**RQ3**: Mapping the **evolution** of privacy categorisations and the **definitions** of the categories.

Additionally, given that privacy relates to ethical issues (e.g. privacy violation, decision-making about the type of information to disclose), this paper explores the extent to which privacy categorisation rely on ethical concepts to improve user's data protection and security. Indeed, privacy not only is a theoretical concept but also an individual disposition, resulting in a concrete behaviour with direct and indirect tangible effects on people's experiences.

In section 2, an overview of privacy management mechanisms and their limitations is presented. Section 3 describes the methodology adopted to carry out the systematic review. Section 4 provides a comprehensive description of Westin's approach and results, followed by a review of the approaches that evolved or departed from it. The latter entails structuring the content of each paper based on its motivation and research questions, methodology and results, positioning it with respect to Westin's approach and our own considerations. Section 5 provides a critical synthesis of the reviewed papers in an attempt to answer to our initial research questions. Finally, section 6 provides concluding remarks and suggests future research directions.

## 2. Privacy and the digital world

Different mechanisms have been proposed to regulate privacy and address the challenges involved in data sharing and protection. Focussing on SNSs, one of the most popular control mechanisms is the 'notice and choice' solution [14]. This solution is based on the idea that users must be notified about the privacy-related implications of information sharing so they can make appropriate informed privacy decisions. However, such a control mechanism was shown to result in little change in users' privacy behaviours [50]. Different studies highlighted not only a lack of knowledge, given that many SNS users have difficulties in managing privacy settings [27, 30], but also a lack of motivation, as users are often unable to fully exploit the control over their data [12]. Additionally, the 'default' solution does not seem to guarantee enough protection. Default settings limits users' sharing tendency only if the users exhibit high privacy concerns [23]. Alternatively, inappropriate defaults (e.g. when the information sharing is heightened) can increase users' privacy concerns and limit their behaviour in sharing information [57, 58].

Managing the plethora of available privacy options can be problematic, as individuals have limited cognitive resources. For example, a limited attention span prevents individuals from carefully evaluating all of the conceivable alternatives and outcomes of their activities. This phenomenon of limited resources is called 'bounded rationality' [43], and it can diminish users' security and privacy protection. In this vein, the inherent uncertainties and ambiguities related to the trade-offs involved in privacy and security decisions can cause user to choose weak settings in order to unlock more functions and gain (apparently) greater value from a particular service. As noted by Camerer et al. [10], decisions involving the disclosure of information or the security of information systems are also susceptible to cognitive and behavioural biases as well as systematic deviations in the judgement and actions of a utility-maximising decision maker. Possible cognitive biases include 'anchoring', which refers to the tendency to consider information as a referent point for a specific situation (e.g. when deciding about posting on SNSs, one may be affected by others' posts); the 'framing effect', which reflects the tendency to make decisions based on how options are presented (e.g. SNS' users are more willing to disclose private information if they are offered stronger privacy controls); and optimism bias and overconfidence, which refer to underestimating the possibility of negative outcomes and overestimating the accuracy of one's judgment, respectively (e.g. users can underestimate the efficacy of antivirus software) [2].

Notably, incomplete and asymmetric information may also create problems related to digital privacy and security [48, 3]. For example, parties that manage mailing lists might sell users' information to other parties without the users' consent. This means that it is difficult for people to understand the risks they are taking by using a specific system or setting, even though they know or acknowledge that their data are being gathered and exploited. The risks may occur when choosing to download an app based on its access to sensitive data, when making a judgment about whether to trust if information should be shared with a website when configuring the browser or cookie settings, when deciding whether to open a link in a document or email or when answering a phone call from an unknown number.

It is also important to point out the 'privacy paradox' phenomenon [36], where there is no or scarce correspondence between privacy-related attitudes and behaviours. This basically reflects the fact that the instruments adopted to measure privacy (e.g. the Internet Users' Information Privacy Concerns scale [32] or the Buchanan's privacy concern and protection scale [9]) have poor predictive power in terms of actual digital behaviour [7, 34]. Although different approaches have been proposed to solve the privacy paradox, such as the 'privacy calculus' model, which focusses on how individuals share information or use privacy settings with regard to benefits and costs [15], criticisms have emerged at both the theoretical and validity levels.

Design techniques like 'nudging' have the potential to improve online privacy and security by guiding user decisions in a subtle manner that does not limit their choices. However, despite its potential to enhance decision-making and minimise errors, this approach is not without limitations. There are concerns regarding the possibility of misaligned judgments, which may result in the alienation of certain users and unintended consequences. The phenomenon in question could result in a transfer of accountability from users, thereby engendering a state of excessive dependence and diminished cognisance with respect to matters of privacy and security. Moreover, the act of nudging may give rise to ethical dilemmas and, in specific contexts, involve the use of manipulative tactics, which could undermine the confidence of users. Consequently, meticulous planning and execution are crucial to maintain the equilibrium between user requirements, ethical deliberations and the probable advantages of nudging. [2]

Privacy categorisation has also been used to support and assist users in making privacy-related choices (e.g. [5, 46].

## 3. Systematic Review Methodology

In this section, we present the methodology used to collect the articles on privacy categorisation.

First, we formulated a list of keywords describing the concept of categorisation we are interested in based on an informal review of the literature. We selected the terms 'segmentation', 'clustering', 'profile', 'profiling' and 'persona' because they have been used to define various forms of categorisation in privacy contexts.

On May 5th, 2022, we began the systematic search of titles and abstracts by combining the chosen keywords with the word 'privacy'. Thus, instead of a single complex string (which would typically be created using Boolean operators and wildcards), five unique search strings were created. This strategy was used to gain a better understanding of how the keyword *privacy* is distributed when associated with segmentation, clustering, profile, profiling and persona.

We used the **5** research strings with three major scientific literature platforms: Scopus, PubMed and Web of Science. Then, all the results were saved. Scopus, PubMed and Web of Science have a number of advantages for conducting a systematic review compared to other resources (e.g. Google Scholar). They have a stringent quality control process, and the papers included in the databases are published in public venues and peer-reviewed journals. Meanwhile, Google Scholar considers a wider array of sources, including theses, preprints and white papers. While these sources can be valuable, their inclusion would require more evaluation to verify the quality and reliability of the research. This was only done for a set of selected articles that were either referenced by other published papers and considered seminal or which concerned our own investigation. Additionally, Scopus, PubMed and Web of Science have advanced search options that allow for more precise, complex queries and the use of controlled parameters. In particular, Scopus and Web of Science, through their citation tracking capabilities, provide information about who cited a particular article, which is useful in assessing the impact and relevance of an article. Further, Scopus, PubMed and Web of Science contain indexed records, making it easier to find articles based on the subject matter.

Based on the Preferred Reporting Items for Systematic Reviews and Meta-Analyses (PRISMA) guidelines [25], all three authors of this systematic review individually undertook the article selection and evaluation process. The aim of the analysis was to identify articles that were pertinent to our systematic review, and it was carried out in three stages. Following the removal of duplicates, articles were first scrutinised by title and then by abstract. Finally, the full text of the selected articles was analysed to determine whether they met the inclusion criteria. Thus, only the papers that passed the title and abstract selection process were read in full. The inclusion criteria were as follows:

- articles written in English;
- focus on privacy categorization;
- proposals of new classifications and labels in the context of the categorization;
- critical analysis, discussion or correlation with Westin's segmentation.

The first two criteria were mandatory for all selected papers, together with at least one of the other two.

| | KEYWORDS | SCOPUS | PUBMED | WEB OF SCIENCE | TOTAL | UNIQUES |
|---|---|---|---|---|---|---|
| | results | **4091** | **363** | **6771** | **11225** | **6193** |
| 1 | **privacy and persona** | 113 | 6 | 63 | 181 | |
| 2 | **privacy and profiling** | 648 | 55 | 2229 | 2930 | |
| 3 | **privacy and profile** | 1134 | 168 | 2229 | 3529 | |
| 4 | **privacy and clustering** | 1823 | 74 | 1988 | 3867 | |
| 5 | **privacy and segmentation** | 373 | 60 | 262 | 695 | |

Table 1: Keyword results for each search engine with details and totals

Following the initial analysis of the 6193 unique papers, we shortlisted 43 papers. After further scrutinising the abstracts of these papers, we selected 13 for a comprehensive text analysis. Of these, the 6 papers listed below were deemed appropriate for inclusion in the systematic review.

- Watson et al. 2015 - Mapping User Preference to Privacy Default Settings
- Dupree et al. 2016 - Privacy Personas: Clustering Users via Attitudes and Behaviors toward Security Practices
- Liu et al. 2016 - Follow My Recommendations: A Personalized Privacy Assistant for Mobile App Permissions
- Wisniewski et al. 2017 - Making Privacy Personal: Profiling Social Network Users to Inform Privacy Education and Nudging
- Dupree et al. 2018 - A Case Study of Using Grounded Analysis as a Requirement Engineering Method: Identifying Personas that Specify Privacy and Security Tool Users
- Toresson et al. 2020 - PISA: A Privacy Impact self-assessment App Using Personas to Relate App Behavior to Risks to smartphone Users

We integrated the search results with 18 additional relevant papers selected based on cross-references (e.g. reference analysis).

In total, we selected, described and analysed **24 papers**, as reported in section 4.

## 4. Literature review

In this section, we review the theories and methodologies of privacy preferences categorization, proceeding in chronological order of publication.

As reported in Fig. 1, the birth of modern privacy profiling began with Westin's studies and evolved through criticism, revision and completely new approaches. The original studies could not account for the plethora of modern problems related to privacy management due to continuously online modern life, which Floridi described as *Onlife* [19], although the founding principles were revised over the years in response to the needs of the digital society.

| ELEMENTS | METHODOLOGY | APPROACH |
|---|---|---|
| Segment | Segmentation | Model driven (may include data analysis, but modelling is prevalent) |
| Cluster | Clustering | Data driven (may include modelling, but data analysis is prevalent) |
| Profile | Profiling | Hybrid (data analysis and modelling are included and are balanced) |
| Personae/Philosophies | Personification | Hybrid (data and model are included, grounded analysis is added) |

Table 2: Elements, methodologies and approaches that constitute *privacy categorisation*

### 4.1. *Westin's methodology (1970-2003): The birth and evolution of Westin's segmentation*

The section examines Westin's significant contribution to the topic of privacy, including a summary of his biography, a look at his foundational writings and a discussion of the creation of his unique segmentation and privacy indices. In addition, the study of Kumaraguru and Cranor (2005), which deeply reviewed Westin's approach, is decribed.

#### 4.1.1. *Westin's short biography*

Alan Furman Westin (1929-2013) was an emeritus professor of public law & government at Columbia University. He was the former publisher of Privacy & American Business and the president of the Center for Social & Legal Research. As a consumer survey expert - mostly for Herris-Equifax in the marketing field - he consulted on more than 100 consumer surveys over his career, covering general privacy, consumer privacy, medical privacy and other privacy-related areas. His well-known privacy segmentation technique is frequently employed in a broad range of applications. Despite the fact that Westin was a prominent historian and professor of privacy legislation, his survey research grew out of his work as a consultant to information-intensive companies [44], and he did not publish it in academic publications. As a result, it has only been subjected to a few in-depth examinations [24].

#### 4.1.2. *Westin's segmentation*

Since its creation, Westin's segmentation has been utilised by academics in a wide range of areas to conduct analyses on privacy. For example, it has been used in psychology, marketing research, computer security and information and communications technology settings. Beyond academia, it is acknowledged that segmentation has also had a significant impact on privacy regulation in the United States [20, 1], where it serves as the foundation for the dominant 'notice and choice' regime, under which consumers are expected to make informed decisions about products and services based on their personal preferences after receiving information about privacy trade-offs. Essentially, the 'notice and choice' model argues that customers will behave as 'privacy pragmatists' and that privacy fundamentalists' preferences are strong enough to affect the marketplace and consumers who are less active [22].

According to the original 1990/1991 work, Westin's privacy segmentation [53], people can be divided into three groups: *Privacy Fundamentalists*, *Privacy Pragmatists* and *Privacy Unconcerned*.

#### *4.1.3. Westin's privacy indices*

Westin created and used multiple *privacy indices*, which evolved over the years. We report some key milestones in the following [53, 54, 24].

The General Privacy Concern Index was established in 1990. Westin utilised a series of four questions to divide respondents into three groups, each of which represented a different degree of privacy concerns [53, 54]:

1. «Whether they are very concerned about threats to their personal privacy today.»
2. «Whether they agree strongly that business organisations excessively seek personal information from consumers.»
3. «Whether they agree strongly that the Federal government has been invading citizens' privacy since Watergate.»
4. «Whether they agree that consumers have lost all control over the distribution of their information.»

The responses to these questions were used to categorise each respondent into one of the following groups based on their level of privacy concern:

- High: 3 or 4 privacy-concerned answers
- Moderate: 2 privacy-concerned answers
- Low: 1 or no privacy-concerned answers

Although based on the questions and degrees of privacy concerns listed above, considering privacy as an ethical value that can be abstracted from the specific domain, Westin proposed indices that have been adapted and renamed based on specific application cases, allowing for their use in the particular case study domain under consideration:

- The Equifax Report on Consumers in the Information Age (1990): *General Privacy Concern Index*
- Harris-Equifax Consumer Privacy Survey (1991): *Consumer Privacy Concern Index*

- Health Information Privacy Survey (1993): *Medical Privacy Concern Index*
- Consumer Privacy Concerns (1993): *Computer Fear Index*
- Equifax-Harris Consumer Privacy Report (1994): *Distrust Index*
- Equifax-Harris Consumer Privacy Report (1996): *Privacy Concern Index*

#### *4.1.4. Latest Westin segmentation categories*

In 2002 [52], Westin provided the most comprehensive summation of the three categories that is available today, known as the *Privacy Segmentation* Index:

- «*Privacy Fundamentalists* (about 25% of the national public): This group believes privacy has an especially high value, rejects the claims that organisations need or are entitled to collect personal information for their business or governmental programmes, thinks more individuals should simply refuse to give out information they are asked for and favours the enactment of strong federal and state laws to secure privacy rights and control organisational discretion.»

- «*Privacy Unconcerned* (about 20%): This group does not understand know what the 'privacy fuss' is all about, supports the benefits of most organisational programs over warnings about privacy abuse, has little issue with supplying their personal information to government authorities or businesses and sees no need to create another government bureaucracy (a 'Federal Big Brother') to protect individual privacy.»

- «*Privacy Pragmatists* (about 55%): This group weighs various business or government programmes calling for personal information, examines the relevance and social propriety of the information sought, wants to understand potential risks to the privacy or security of their information, seeks to confirm whether fair information practices are observed and then makes decisions about the specific information-related activities of industries or companies. Pragmatists favor voluntary standards and consumer choice over legislation and government enforcement. However, they will back the legislation if they think that not enough is being done volunarily.»

Based on the summary table in Kumaraguru's work [24], from 1990 to 2003 the application of Westin's segmentation was based on the emergence of three privacy concern groups (High, Medium, Low). Those groups were the basis for the development of the various indices (e.g. Consumer Privacy Concern Index, Medical Sensitivity Index, Distrust Index). The group in the middle (Medium/Pragmatists) was the largest, and this uneven population was the basis for works that critiqued, reworked and expanded Westin's work in the following years.

### *4.2. Kumaraguru and Cranor (2005)*

Kumaraguru and Cranor [24] presented a report to help researchers better understand Westin's work. They showed that most of Westin's indices cannot be directly compared, and thus the procedure used by Westin to develop the indices (e.g. 1990 [54] and 1996 [55] studies) was incorrect. Specifically, the indices utilised in the different studies did not use the same criteria (questions), and because the options (answers) used for obtaining the indexes differed across studies, it is not possible to compare them. Moreover, Westin did not construct or offer procedures or comparison criteria [53] to support a more direct comparison.

They also proposed a summary of the different aspects that Westin used for deriving the privacy indices:

- General Privacy Concern Index (1990): Whether they are very concerned about threats to their personal privacy today. Whether they agree strongly that business organisations excessively seek personal information from consumers. Whether they agree strongly that the Federal government has been invading citizens' privacy since Watergate. Whether they agree that consumers have lost all control over the distribution of their information.

- Consumer Privacy Concern Index (1991): Agreement with the statements: Consumers have lost all control over how personal information about them is circulated and used by companies.
  My privacy rights as a consumer in credit reporting are adequately protected today by law and business practices.

- Medical Privacy Concern Index (1993): Whether they have ever used the services of a psychologist, psychiatrist or other mental health professional. Do you believe your personal information has been disclosed? There were four other questions that all related to medical information.

- Computer Fear Index (1993): If privacy is to be preserved, the use of computers must be sharply restricted in the future. Concern level in usage of computers in medical services (patient billing, accounting).

- Distrust Index (1994): Technology has almost gotten out of control. Government can generally be trusted to look after our interests. The way one votes has no effect on what the government does. In general, business helps more than harms.

- Privacy Segmentation and Core Privacy Orientation Index (1995-2003): Consumers have lost all control over how personal information is collected and used by companies. Most businesses handle the personal information they collect about on in a proper and confidential way. Existing laws and organisational practices provide a reasonable level of protection for consumer privacy today.

### 4.3. *Evolution, Critics, and Departures from Westin's Segmentation*

Below, we list the works we found during our research that took their cues from Westin, critiquing it, extending it or using it in contexts other than the original one, from mobile applications to health. These works are reported in chronological order of appearance. Notably, with the exception of *Hoofnagle et al. (2014)* [22], who theoretically analyzed Westin's work, all of the reviewed research works pertain to the digital world.

The analysis extracted the following data:

- motivation;
- research questions;
- design and methodology, including sample characteristics, instruments and statistical analyses;
- results, including type of privacy categories if any; criticisms and/or advancements of Westin's approach;
- strengths and weaknesses of the study.

**Sheehan (2002)** [42]

*Motivations*
Starting with the analysis of Westin's marketing-based research, the paper aims to characterise online users' behaviour.

*Research questions*
Examining the different types of Internet users' online privacy concerns, how these different types of privacy concerns relate to each other and how they affect Internet users' behaviour.

*Methodology*
A total of 889 Internet users were enrolled. The participants completed a survey to indicate their concerns about their privacy in 15 different situations (e.g. sharing their name, address and phone number online) from the perspective of a personal (as opposed to commercial) user of the Internet. A seven-point bipolar scale was used, ranging from 1 (not at all bothered) to 7 (very concerned). The survey also asked participants about their demographic information (e.g. age, gender, education and income). A 'total concern' score (ranging from 15 to 105) was created by summing each of the concern scores for the 15 situations: the higher the score, the higher the participant's concern with privacy regardless of the situation. The total concern score was used to categorise participants into three groups, which corresponded to Westin's segmentation. Then, based on the distribution of the total concern score, a fourth group was identified by dividing the original Westin's 'pragmatist' segment into two groups. A series of analysis of variance (ANOVA) and chi-squared tests were performed to analyse inter-group differences in demographics and computer usage and actions.

*Results*
Four groups were defined: 1) Unconcerned Internet users (score of 30 or less; older than average and bachelor's education or less) - minimal concern with online privacy and provided highly accurate information for web sites; 2) Circumspect Internet users (score between 31 and 60; younger than average and lower levels of education) - minimal concern with online privacy overall, similar to the unconcerned Internet users, although they sometimes provided incomplete information in their registrations; 3) Wary Internet users (score between 61 to 89; younger and better educated) - moderate level of concern with online privacy in many situations and high concern in several situations; occasional complaining and incomplete information provided at the moment of registration; 4) Alarmed Internet users (score above 91; older with higher levels of education) - highly concerned about online privacy, high level of complaining and rarely registered for web sites;

*Criticisms and proposed advancement to Westin's approach*
Westin's segmentation is excessively comprehensive and fails to highlight the complexity of users' online privacy concerns. Westin's tripartite segmentation is too limited and therefore was extended to four distinct typologies, given that the 'pragmatists' can be divided into two different groups.

*Our Considerations*
The taxonomy of privacy concerns is a valuable instrument for comprehending individuals' perceptions of online privacy. Using a large sample, the study additionally explored the determinants of individuals' privacy concerns, including but not limited to demographic characteristics, educational background and prior Internet usage. The study was the first to examine privacy concerns into the social context, arguing that users' privacy concerns are influenced by interpersonal relationships, cultural background and social norms. Regarding limitations, first the study relied on data from the United States, which are characterised by a specific society and culture. Second, the paper did not examine the impacts of the different forms of privacy concerns on Internet-related behaviour. Third, the construction of the four groups was not validated by standardised statistics methods.

**Berendt et al. (2005)** [8]

*Motivations*
The paper analysed consumers' privacy behaviour in e-commerce contexts.

*Research Questions*

Examining the extent to which users' stated privacy preferences align with their actual behaviour when shopping online and defining which factors affect the discrepancy between their stated preferences and actual behaviour.

*Methodology*
A total of 171 online shoppers were enrolled. A combination of surveys, interviews and focus groups was used. The survey asked participants about their privacy concerns, online behaviour and demographic information. The interviews and focus groups allowed participants to discuss their privacy concerns in more detail. Subsequently, the participants were involved in a simulation where they purchased cameras and clothing online, which were discounted by 60% off of local shop pricing. An anthropomorphic shopping bot helped the participants with the purchase. The participants who chose to purchase had to pay for the items. Based on that, a Personal Consumer Information Cost (PCIC) index was developed, taking into account the validity and relevance of each response in the sales environment as well as the difficulty of responding it. A PCIC index of 'zero' suggests the user can answer the question truthfully. A high PCIC index indicates consumers are hesitant to provide the related information. The PCIC index was associated with legitimacy and relevance and somewhat correlated with difficulty, based on a regression analysis. A cluster analysis was also performed.

*Results*
A significant discrepancy was found between users' stated privacy preferences and their actual behaviour when shopping online. The factors that affected this discrepancy were as follows: 1) The perceived benefits: Users may be more willing to provide personal information if they believe that they will receive some benefit in return, such as a discount or a personalised shopping experience. 2) The perceived risks: Users may be less willing to provide personal information if they believe that their privacy is at risk. 3) The perceived ease: Users may be more willing to provide personal information if it is easy to do so. 4) The perceived importance of privacy: Users may be more willing to provide personal information if they do not believe that privacy is important. The clusters were defined as follows: *Privacy Fundamentalists, Profiling Averse, Marginally Concerned and Identity Concerned*. In particular, privacy fundamentalists and marginally concerned individuals are worried about giving personal information, such as their name, email or postal address, whereas profiling averse users are more concerned about sharing personal information about their interests, hobbies and health condition.

*Criticisms and Proposed Advancements to Westin's Approach*
This paper did not directly criticise Westin's work on privacy concerns. However, it suggested that among Westin's privacy concern indices, the highest concern is for the privacy of personal information. Additionally, this study proposed a more nuanced view of online privacy concerns, suggesting that there is a continuum of privacy concerns and that people's concerns can change over time. Further, Berendt argued that people's privacy concerns develop in both private and social dimensions.

*Our Considerations*
The study highlighted the discrepancy between intention and actual behaviour and underlined the key role of the social context in shaping privacy concerns. The need for new approaches to protect online privacy was also emphasised. Regarding limitations, first, the sample size was relatively small and was unbalanced in terms of age, education and culture. Second, the study was based on self-reported and self-disclosed data. Participants may not always be honest about their privacy concerns or their online behaviour. Third, the statistical approach was not fully explained, with possible confounding effects due to intervening variables, such as age, gender, education and technical proficiency.

**Consolvo et al. (2005)** [13]

*Motivations*
The paper aimed to characterise the decision-making process related to sharing personal information in a social relations context to improve the design of future location-enhanced applications and services.

*Research Questions*
Identifying the factors that influence people's decisions to disclose their location, clarifying how people use location disclosure to maintain social relationships and defining the privacy implications of location disclosure.

*Methodology*
A total of 16 participants were enrolled. The study used a combination of methods in three phases, including a questionnaire about the users' social networks and how they expected to utilise location-enhanced computing (phase 1). In addition, experience sampling was performed using a mobile application to determine users' intention to disclose their location information based on the hypothetical requests from people on the buddy lists created in phase 1 (phase 2). Interviews were conducted to gather the participants' thoughts about their experiences (phase 3). Westin's segmentation model was used in phase 1 to determine the groups.

*Results*
The following factors were found to affect people's decision to disclose their location: 1) The relationship between the discloser and the recipient: People are more likely to disclose their location to people they know and trust. 2) The context of the disclosure: People are more likely to disclose their location in certain contexts, such as when they are meeting up with friends or when they are travelling. 3) The perceived benefits of disclosure: People are more likely to disclose their location if they believe that it will have benefits, such as making it easier to meet up with friends or to stay safe. 4) The perceived risks of disclo-sure: People are less likely to disclose their location if they believe that entails risks, such as being tracked by someone they do not know or being targeted by advertising. Additionally, pragamtists were found to share their location according to the context and outcomes. However, Westin's privacy classification was not a good predictor of how users would respond to location requests from social relations. In addition, the results showed that the participants either revealed the most helpful (but not necessarily the most thorough) information about their location or did not disclose it at all. User location and activity were found to be of lesser importance.

*Criticisms and Proposed Advancements to Westin's Approach*
This paper did not directly criticise Westin's work on privacy concerns. However, it suggested that Westin's typology of privacy concerns may not be entirely accurate regarding users' intention to share location-based data for the pragmatist group. The study also focused on the social context of privacy concerns, highlighting that the social relationship between the sharer and other participants matters and arguing that the technology used for location disclosure has changed the way people think about privacy. Further, the key role of the user's experience of location disclosure was underlined in designing privacy-protecting technologies.

*Our Considerations*
The study provided a valuable contribution to the understanding of location disclosure. The results suggested that there is a need for more research on the factors that influence people's decisions to disclose their location to others, on the ways that people use location disclosure to maintain social relationships and on the privacy implications of location disclosure. However the study is based on a very small sample size (16 participants), although the data collected for each participant were relevant to the aims of the study. In addition, the study was based on self-reported data, and the participants were asked to report their thoughts and feelings about location disclosure. Finally, the statistical approach was not clearly described in the paper.

**Hoofnagle et al. (2014)** [22] and **Urban et al. (2014)** [47]

*Motivations*
These papers theoretically criticised Westin's 'homo economicus' categorisation and proposed disentangling the economic dimension of privacy to support a political discourse on privacy.

*Research Questions*
While the research questions were not explicitly stated, the authors aimed to clarify the extent to which Westin's typology of privacy concerns maps onto economic theories of privacy and consequently how the 'homo economicus' model accounts for the ways in which people make decisions about their privacy. Further, they sought to identify the strengths and weaknesses of Westin's approach to privacy, to clarify how Westin's model could be used to improve our understanding of privacy and to determine the implications of Westin's approach for privacy policy.

*Methodology*

The study comprised a theoretical analysis and critic to the Westin's work and an empirical experiment involving 2203 subjects in two rounds completed in 2009 (1000 subjects) and 2012 (1203 subjects). The participants were presented with specific information privacy propositions available in the marketplace with the aim to understand their preferences and control levels. Scenario-based testing was employed to elicit privacy concerns related to new services. Westin's three screening questions were used to divide the respondents into three groups: pragmatists, fundamentalists and the unconcerned. Then, consumers' familiarity with and opinions on a wide range of topics were examined, which evolved along with the market. Finally, the customers were mapped to Westin's privacy segmentation to evaluate its efficacy for a few of the queries.

*Results*

Westin's homo economicus privacy model was found to be a useful tool for understanding privacy, but only if used in conjunction with other models. Westin's privacy segmentation model inaccurately labelled a broad group of American consumers as 'pragmatists' without establishing whether they actually engaged in the kind of deliberations that define pragmatism. Empirical research revealed that many consumers have fundamental misunderstandings about business practices, privacy protections and restrictions on the use of data. These misunderstandings cause them to expect more protection than what is currently offered. When presented with specific information about the privacy propositions available in the marketplace, most consumers prefer more control than they currently have. Consumers' misunderstandings distort the market for privacy because they lead consumers to believe they do not need to negotiate for privacy protections. Many individuals' decisions are deeply misinformed about business practices and legal protections. Westin's pragmatists were found to understand less than either the fundamentalists or the unconcerned. Contrary to Westin's description, when presented with real-world scenarios reflecting privacy concerns about new services, the pragmatists joined the fundamentalists in rejecting information-intensive service options.

*Criticisms and Proposed Advancements of Westin's Approach*

Although the study recognised that Westin' model is relatively simple and easy to understand, it criticised the model for inaccurately labelling a large group of consumers as pragmatists without verifying their actual deliberative behaviours. Morever, the study contended that the model overestimates consumers' understanding of business practices and privacy protections, leading to a false sense of security. Furthermore, it criticised the model for placing the burden on consumers to negotiate for privacy protections in the marketplace. Finally, the study argued that Westin's model makes consumers' behaviour the cause for the spread of privacy-invasive services, deflecting the focus away from necessary changes in the structure of the marketplaces.

*Our Considerations*
The studies were based on a theoretical analysis of Westin's work and it's empirical validation. However, an empirical study using different models and comparing them with the one proposed by Westin would be needed to test the validity of the paper's arguments.

**Lin et al. (2014)** [26]

*Motivations*
In order to help and support the user in setting app permissions, this work aimed to show that it is possible to identify a small number of privacy profiles that reflect diverse permission preferences.

*Research Questions*
While the research questions were not explicitly stated, the study aimed to simplify mobile app privacy settings management, to address the feasibility of categorising users into privacy profiles and to clarify the influence of the purpose of an app's request on users' comfort with permissions.

*Methodology*
The study used static code analysis to identify the purposes of app permissions. A user survey was then conducted using Amazon Mechanical Turk (AMT), where the participants rated their comfort levels regarding these app permissions based on their purpose. Survey tasks were structured around specific app-permission-purpose sets identified through the code analysis. The survey included 1,200 tasks, covering 837 mobile apps, with the aim to recruit 20 unique participants per task. The final data set consisted of 21,657 responses from 725 AMT workers. The paper employed hierarchical clustering with an agglomerative approach to cluster mobile app privacy preferences. The selection criteria included evaluating dendrogram structures and internal measures, such as connectivity, silhouette width and the Dunn Index.

*Results*
Significant differences were found in the participants' comfort levels regarding various app permissions. Participants were most comfortable with apps using location information for internal functionality and social networking services (SNS) using location information for sharing. Discomfort was noted with targeted advertising libraries accessing private information, SNS libraries accessing phone IDs and contact lists and mobile analytic libraries accessing location and phone state. A high variance in privacy preferences was found, indicating that a one-size-fits-all privacy setting would be insufficient.

Using the Canberra distance and average linkage method, the study identified four clusters of users based on their privacy preferences:

- *privacy conservatives* (11.90% of participants; lack of comfort granting permissions; uncomfortable with mobile apps asking to access phone ID, contact list or SMS functionality);
- *unconcerned* (23.34% of participants; high level of comfort disclosing sensitive personal data, with the exception of granting SNS libraries access to the Get_Accounts permission (e.g. information linked to Facebook, Google+, Youtube; in general they are younger and have lower levels of education);
- *fence-sitters* (approximately 50% of participants; in between the extremes, being quite comfortable disclosing sensitive personal data; similar to pragmatists);
- *advanced users* (17.95% of participants; highly nuanced understanding of which usage scenarios they should be concerned about, e.g. they dislike targeted ads and mobile analytic libraries but agree to disclose coarse location; in general, they are older and with have a higher level of education).

Demographics, such as gender and age, did not significantly impact cluster assignments, but education level showed some correlation. Privacy profiles served as initial settings that users could personalise according to their preferences.

*Criticisms and Proposed Advancements of Westin's Approach*
Westin's privacy indices were used as a reference and to support the study's findings. Thus, the paper acknowledged Westin's findings and drew a parallel by identifying similar groups or clusters of users based on their privacy preferences. Further, the paper highlighted the diversity of users' privacy preferences and the need for personalised privacy settings. While Westin's work provided valuable insights on user privacy attitudes, this study took a more data-driven approach, using clustering techniques and crowd-sourcing to identify distinct privacy profiles and proposing default settings tailored to users' preferences.

*Our Considerations*
This study contributed to the field by quantitatively linking app privacy behaviours to users' privacy preferences, identifying distinct privacy profiles and proposing automated privacy settings. The study's large-scale data collection and systematic statistical approach provided valuable insights on mobile app users' diverse privacy preferences and offered a foundation for improving privacy controls. The study focused on free apps from the Google Play Store, which limits the generalisability of the findings, as paid apps may elicit different privacy-related behaviours. As acknowledge by the authors, the coarse classification used to determine why sensitive resources are requested overlooks finer distinctions, and the reliance on static analysis may not have captured dynamic privacy behaviours.

**Liu et al. (2014)** [29]

*Motivations*
The papers aimed to define personalised classifiers by identifying privacy profiles to reduce the burden on users while giving them better control over app permissions.

*Research Questions*
Understanding people's privacy preferences with respect to permissions in different mobile apps by utilising personalised classifiers and privacy profiles

*Methodology*
The methodology of the study consisted of the following steps:
1) Data Collection: Data were gathered over a 10-day period from 4.8 million users using the LBE Privacy Guard app, an Android application that allows users to manage app permissions.
2) Data Pre-processing: Ths focused on 'representative users' and 'representative apps' for a more robust analysis. Users who installed at least 20 apps and manually selected at least one 'Deny' or 'Ask' permission were chosen. Apps with at least one permission request, having at least 10 users and available on the Google Play store during the data collection period were selected.
3) Data Analysis: User patterns and preferences regarding app permissions were identified. The aim was to predict likely user responses to permission requests based on privacy profiles.
4) Model Evaluation: The predictive models' effectiveness in anticipating user preferences was assessed, aiming to simplify permission control for users while maintaining their agency.

*Results*
Three to six privacy profiles were created. Each user was modeled as a 12-dimensional vector of app-permission decisions (1 = allow; -1 = deny), with profiles relying on single permissions (5) and permission pairs (5) with the highest discriminating scores. The high accuracy rate achieved, exceeding 87%, demonstrated the effectiveness of this approach in capturing user preferences. Overall, the results showed the potential of privacy profiles to simplify app permission decisions and achieve high accuracy in predicting user preferences. In general, the results indicated that it is feasible to dramatically minimise the user burden while still allowing consumers to have more control over their mobile app permissions. The study demonstrated that simple tailored classifiers might be developed to anticipate a user's app permission choices.

*Criticisms and Proposed Advancements of Westin's Approach*
With respect to Westin, the study proposed a new approach to simplifying privacy decisions using personalised classifiers and privacy profiles.

*Our Considerations*

One of the notable strengths of the study is that it relied on an innovative theoretical and methodological approach. By leveraging personalised classifiers and privacy profiles, the study offered a way to predict user app permission decisions based on individual preferences. Additionally, the research highlighted the potential to significantly reduce the user burden. Through the utilisation of privacy profiles, users can align their preferences with like-minded individuals, simplifying the decision-making process and enhancing usability. Regarding limitations, first, the study relied on the LBE dataset where users should be technically proficient in order to root the Android device and install the LBE Privacy Guard system. This may not fully represent the diverse range of app users and their privacy preferences. Second, the metrics introduced were too subjective. Further studies should be conducted to ascertain their validity and real-world applicability.

**Woodruff et al. (2014)** [61]

*Motivations*
This paper sought to understand the relationships between Westin's Privacy Segmentation Index and the gap between privacy attitudes and behaviours.

*Research Questions*
Addressing the correlation between Westin's Privacy Segmentation Index, behavioural intentions, attitudes and consequences of privacy behaviours, especially in response to specific privacy scenarios and outcomes and clarifying whether Westin's Privacy Segmentation Index can be improved or supplemented by other variables, such as personality traits and demographics, to predict responses to privacy scenarios and outcomes.

*Methodology*
A total of 884 participants were enrolled in this study, which was based on a two-phase approach involving AMT and Google consumer surveys. In the first phase, a survey was administered to capture general privacy attitudes using the Westin Privacy Segmentation Index as well as other scales related to privacy concerns, participants' degree of direct and/or indirect experience with the misuse of personal information and personality traits using scales from the psychology literature. Specifically, the following were used: the Ten Item Personality Inventory, locus of control, moral foundation theory, general disclosiveness (amount, depth and honesty subscales), generalised self-efficacy, Stimulating-Instrumental Risk Inventory, ambiguity tolerance, hyperbolic discounting and Cognitive Reflection test. In the second phase, participants were asked to imagine themselves in three out of 20 randomly chosen privacy scenarios and to assess their own attitudes and behavioural intentions using a Likert scale. The participants were also presented with outcomes associated with the scenarios and asked to evaluate their likelihood of disclosure. Then, correlation/regression analyses and one-way ANOVAs were performed.

*Results*
Regarding the Westin Privacy Segmentation Index, the distribution of responses revealed that the majority of participants fell into the fundamentalist category, followed by pragmatists and the unconcerned. Demographic variables and personality traits did not significantly predict Westin's categories. In terms of scenario responses, there were no significant differences between Westin's categories, indicating a lack of association. In general, there was no attitude–behaviour dichotomy or attitude–consequence dichotomy, consistent with the individual items or derived categories of the Westin Privacy Segmentation Index.

*Criticisms and Proposed Advancements of Westin's Approach*
The study questioned the effectiveness of Westin's categories in predicting privacy-related behaviours. Alternative instruments and segmentation approaches were suggested for further research, considering context-specific factors and deep-seated preferences for privacy. The study also highlighted the need to explore the trade-off between clustering preferences and context-specific decisions.

*Our Considerations*
This paper involved a comprehensive exploration of the relationship between the Westin Privacy Segmentation Index and participants' responses to hypothetical scenarios and outcomes. Various factors were considered, such as personality traits, demographics, situational variables and the use of statistical techniques to assess the predictive power of the index. The paper also provided insights on the limitations of the Westin Privacy Segmentation Index and highlighted the need for further research and alternative approaches to better understand privacy behaviours. While the authors highlight the limits of the Westin segmentation, the lack of predictive power they report in the discussion could also be influenced by the AMT sampling and the use of self-report instruments.

**Watson et al. (2015)** [51]

*Motivations*
This papers aimed to explore the complexity of managing online privacy and the challenges users face in configuring and adjusting privacy settings.

*Research Questions*
Investigating whether default privacy settings on social network sites (Facebook) can be customised to better match the preferences of users.

*Methodology*
A survey of 184 Facebook users (age range 19–66, mean age of 31.4, male = 104) was conducted out to gather data on privacy profile preferences and reactions to changes in audience settings. Participants were recruited using the AMT platform. The questionnaire consisted of three components: demographics, general privacy attitude questions and specific questions about privacy preferences for 29 profile items. The participants were asked to indicate their preferred sharing audience for each profile item and their attitudes toward alternate audience disclosures. The survey data were used to compute fit scores representing the alignment between applied policies and user preferences. Thus, an optimal policy based on the reported preferences of a training sample was generated and compared with different default policies, a completely restrictive policy (Restrictive), the participants' preferred audience (Mode) and the permissive Facebook default settings. Based on the usage and general privacy attitudes, three privacy segmentation models were derived: the Westin/Harris' model (pragmatist, fundamentalist and unconcerned) and Buchanan's and Facebook Intensity Index models, based on which the participants were divided into low, average and high according to the standard deviation from the means (the average group ranged from -1 to +1 standard deviation, whereas the low and high groups were below -1 and above +1 standard deviations from the means, respectively). Then, these models were used to determine whether multiple canonical policies improved the default settings

*Result*
The participants demonstrated a preference for sharing profile information with friends only, especially for sensitive items, while their preferences varied for less sensitive items. The participants' characterisations of disclosure desirability for different audience choices showed that more restrictive audiences were generally viewed as neutral, while more permissive audiences were moderately undesirable. The results demonstrated that the calculated policies, including the optimal one, had different characteristics. The optimal policy tended to prioritise more restrictive settings based on the participants' preferences, while the mode policy reflected popular choices. The fit scores analysis revealed significant differences between the policies. The optimal policy had the highest fit scores, indicating a better alignment with participants' preferences. In contrast, the Facebook default policy had the lowest fit scores, suggesting a mismatch with user preferences. Interestingly, the mode and restrictive policies did not significantly differ in terms of fit scores, indicating that both approaches were similarly effective in representing user preferences.

*Criticisms and Proposed Advancements of Westin's Approach*
Criticisms included the limitations of privacy attitude segmentation models, such as their inability to capture contextual privacy attitudes on social media platforms like Facebook. The simplicity of the segmentation techniques used in the index scores was also a limitation. Proposed advancements included exploring more sophisticated segmentation models, potentially using supervised machine learning techniques and larger training sets. The paper suggested that the current approach may not adequately capture the diverse privacy preferences within online social networks. Additionally, the limitations of default policies and the burden of configuration were highlighted. The paper proposed gathering additional user information to generate more personalised and privacy-preserving default settings. Additional research was recommended to investigate the trade-off between effort and the configuration burden as well as to explore novel methods for minimising the effort required to manage online privacy.

*Our Considerations*
The study presented a comprehensive analysis of privacy preferences and default policies in online social networks. It highlighted the potential to improve default privacy settings to better align with user preferences and enhance privacy management on social network sites. Regarding limitations, the sample size was relatively small. Additionally, as also discussed by the authors, the reliance on self-reported privacy preferences and attitudes could have introduced biases and discrepancies between reported behaviour and actual user actions. The segmentation models used to categorise privacy attitudes have been questioned for their limited ability to capture the complex and contextual nature of privacy preferences in online social networks. Further, the authors acknowledged that the default policies proposed may err on the side of being more restrictive, potentially hindering social interactions and reducing the value of the platform.

**Liu al. (2016)** [28]

*Motivations*
The paper aimed to develop a personalised privacy assistant to help users manage privacy preferences.

*Research Questions*
Exploring the effectiveness of a personalised privacy assistant (PPA) in providing suitable recommendations for mobile app permission settings to users, investigating the extent to which users adopt the recommendations offered by the PPA, examining how users engage with the privacy nudges presented by the PPA, exploring the frequency with which users modify the permission settings initially suggested by the PPA and studying users' perceptions of the usefulness and usability of the PPA and its recommendations.

*Methodology*
The methodology involved conducting field studies with Android users who had rooted devices and used them for more than one month. The study initially had 131 participants, but after some were excluded, the final sample size was 72 participants. Participants were selected from online communities and had to meet certain criteria (e.g. using a rooted Android phone with a data plan, 18 years or older). Data were collected through an app that captured participants' permission settings and app usage. The collected data included permission settings, app categories and purpose information. The data analysis involved building privacy profiles using hierarchical clustering and training a classifier for personalised recommendations. Hierarchical clustering was used to build privacy profiles based on aggregated preferences. A scalable support vector machine classifier (LibLinear) was trained using the collected permission settings to generate personalised recommendations. Logistic regression models were applied to analyse the impact of different factors on users' permission settings. A down-sampling analysis was conducted to assess the effectiveness of the profiles with different data set sizes.

*Results*

Seven privacy profiles were created. In light of the profiles identified by Lin et al. (2014) [26], the results showed that profiles 1, 2, 5, 6 and 7 aligned with 'fence-sitter' and 'advanced user' profiles; profile 3 corresponded to the 'unconcerned' profile; profile 4 corresponded to the 'conservative' profile. The profiles were then used to evaluate the effectiveness and usability of the profile-based PPA for mobile app permissions. Seventy-two users (different from the previous field study) were included (49 in the treatment group and 23 in the control group). The results showed that 78.7% of the recommendations made by the PPA were accepted, whereas only 5.1% of the recommendations were revised by participants in the treatment group as compared to the control group. In addition, the treatment group converged faster on their settings and were also satisfied with the recommendations and the PPA. In addition, the participants felt comfortable with the recommendations and reported improved privacy. The PPA was perceived as useful, particularly with regard to app monitoring and usability. The recommendations were found to be helpful in configuration and made decision-making easier for the participants.

*Criticisms and Proposed Advancements of Westin's Approach*

The study showed similarities to Westin's approach by focusing on understanding and addressing individuals' privacy preferences. The PPA aimed to assist users in configuring their mobile app permissions based on their unique privacy preferences, aligning with Westin's segmentation approach, which categorises individuals into groups based on their privacy attitudes. Additionally, the emphasis of the study on personalised assistance and tailored recommendations suggested a potential advancement by providing more accurate and fine-grained privacy recommendations. In general, this study reflected a practical application of Westin's broader goals for understanding and accommodating individual privacy preferences.

*Our Considerations*

The study involved field studies and deployment of the PPA on participants' smartphones, increasing the ecological validity of the findings. The proposed assistant learned privacy profiles and provided tailored recommendations, effectively assisting users in configuring app permissions based on their preferences. The paper collected comprehensive permission data and aggregated the data along different dimensions, resulting in an in-depth analysis. However, the recruitment of rooted Android device users may have limited the generalisability of the findings to a broader population. In addition, the relatively short study duration may have limited the assessment of long-term effectiveness and user preference stability. The paper did not explicitly compare the PPA to other privacy management tools or approaches. The participants suggested enhance-ments related to the timing and modality of privacy nudges, providing more information about the impact of permissions and incorporating purpose-centric controls for permissions. Age, gender and educational factors were not addressed in the statistical analyses.

**Wisniewski et al. (2017)** [59]

*Motivations*
This paper examined why social networks users do not fully exploit privacy controls but instead apply privacy strategies related to their privacy awareness.

*Research question*
Profiling Facebook users both in terms of feature awareness and privacy behaviour and exploring the relationships between users' privacy awareness and behaviour in order to understand whether users' privacy management strategies are affected primarily by conscious behaviours or by their limited knowledge of the available privacy controls.

*Methodology*
A total of 308 Facebook users were enrolled in this study. Both feature awareness and privacy behaviour were measured through a self-report questionnaire. The questionnaire was focused on the settings adopted to manage interpersonal privacy boundaries (e.g. I did not provide this information to Facebook; How often have you done the following to modify posts on your News Feed?) and the proficiency related to a specific interface feature or functionality useful for a task (e.g. I vaguely recall seeing this item). First, confirmatory factor analyses were performed to determine the dimensional structure of both feature awareness and privacy behaviour. Then, a structural equation model (SEM) was used to test the relationship between feature awareness and privacy behaviour. Finally, mixture factor analysis was applied to the confirmed factors to cluster participants based on their varying dimensions of feature awareness and privacy behaviours. Based on a mixture factor analysis, each participant was assigned to one of K classes, minimising the residual difference between the observed and predicted factor scores for each participant. Finally, the bi-directional overlap in class membership between privacy management strategies privacy awareness was examined.

*Results*
Six class solutions for privacy behaviour management strategies (management profiles) and six awareness profiles for privacy proficiency were obtained. Management profiles:
1. privacy maximisers - higher levels of privacy across the most of privacy features;
2. self-censors - infrequent use of privacy features and settings, but high withholding of personal information;
3. time savers/consumers - similar to privacy minimalists, but passive consump-

tion of Facebook updates, such as restriction of chat availability;
4. privacy balancers - moderate levels of privacy management behaviours;
5. selective sharers - advanced privacy settings, such as the creation of friend lists and posting content selectively to these groups;
6. privacy minimalists - fewer privacy strategies, such as limiting Facebook profile by default).

Proficiency profiles varying in degree, from the most basic to the highest level: 1. novices; 2. near-novices; 3. mostly novices; 4. some expertise; 5. near-experts; 6. experts. In general, there was some overlap between the privacy management profiles and privacy proficiency profiles. Privacy maximisers were experts or near-experts, self-censors and time savers/consumers exhibited intermediate levels of proficiency and privacy balancers showed higher or intermediate levels of expertise or were complete novices. Meanwhile, selective sharers showed higher levels of expertise, while privacy minimalists ranged from mostly novices to complete novices.

*Criticisms and Proposed Advancement of Westin's Approach*
Based on existing critiques (Woodruff et al., 2014) of Westin's approach, this study explored behaviours related to both informational and interactional privacy boundaries. Specifically, unlike Westin's coarse categorisation, six privacy management strategies were empirically derived from self-reported data, highlighting privacy behaviours that are mostly common and infrequent. Users were found to exhibit distinctly different behavioural patterns rather than more or fewer privacy behaviours. Importantly, the study was not limited to privacy behaviour but also considered privacy awareness, highlighting the fact that users first learn the most basic privacy features and then the more advanced ones. In general, privacy awareness was found to predict privacy behaviour.

*Our Considerations*
The study adopted a sound approach to understand users' privacy management strategies and privacy awareness. The relationships between these two privacy-related aspects was examined by combining advanced statistical techniques, including confirmatory factor and mixture factor analyses. Additionally, the study clearly showed the multi-dimensional structure of privacy, both in terms of behaviour and awareness, highlighting the key role of awareness in privacy-related behaviour. Finally, the implications of the results for privacy education and nudging were discussed, along with specific recommendations for improving these interventions. Regarding limitations, the sample size of 308 Facebook users was relatively small and limits the generalisability of the findings. In addition, the data were not corrected for age, gender and educational level. The study relied on self-reported data tied to Facebook characteristics, thus the results are not generalisable to all social-networks (e.g. LinkedIn, Instagram) or to online experiences outside of social networking in general.

**Dupree et al. (2016-2018)** [17, 18]

*Motivations*
The first study (2016) aimed to define users' categorisation based on their attitudes and behaviours toward security practices. The second study (2018) sought to address several key aspects related to requirements engineering and the development of a PPA for mobile app permissions.

*Research Questions*
The first study (2016) examined the distinctions between user clusters and the categories established by Westin. In addition, it assessed the coherence and consistency of identified user clusters across different participant samples and explored the design implications of user clusters for the development of security and privacy technologies. The second study (2018) explored the importance of user-space identification and categorisation, the creation and application of user-space-covering personas, the use of grounded analysis in producing a specification as a grounded theory and the significance of privacy and security features in computer-based systems.

*Methodology*
In the first study (2016), three rounds of sampling with different tests were conducted, with a total of more than 200 participants. The main data used in both papers to understand users' privacy and security concerns were collected from 32 university-educated participants aged 22 to 35, primarily from a population of non-computer science graduate students. An additional set of 13 participants was interviewed remotely via Skype. An agglomerative clustering approach was used in this study, where the participants were clustered by creating a weighted graph that visualised connections between them. Edge weights in the graph represented the number of shared traits between participants. Similarity between participants was measured using dot-product calculations. Traits shared by too many clusters were eliminated, and the clustering was refined using a procedure inspired by latent semantic analysis, a textual analysis technique. The methodology used in the second study (2018) was based on grounded analysis, which involves iterative coding and categorisation of data to develop a comprehensive understanding of user behaviour and characteristics within the privacy and security tool user space. The study utilised a case study approach to validate the effectiveness of the method, and it involved conducting interviews. The study employed a two-step categorisation process to create personas. First, users were analyzed using Westin's categorisation (pragmatist, fundamentalist and unconcerned). Through this procedure, two dimensions emerged, which were used to describe the participants, knowledge and motivation, especially with respect to the pragmatist category. Then, in the second categorisation, similarities in users' *quotations* were combined with the grading of participants based on knowledge and motivation, leading to the development of five personas

*Results*
In the first study (2016), five clusters were identified based on security and privacy behaviours, including Fundamentalists, Lazy Experts, Technicians, Ama-teurs and the Marginally Concerned. The findings suggested that the five-cluster solution provides a more nuanced understanding of user categorisation compared to traditional approaches, with implications for designing effective security and privacy tools. In the second paper (2018), five personas were created as a result of the categorisation process. The personas represented different levels of knowledge and motivation toward privacy and security:

- *Mark, marginally aware* (low knowledge and motivation);
- *Robert, fundamentalist* (high knowledge and motivation);
- *Allison, struggling amateur* (medium knowledge and motivation);
- *Patricia, technician* (medium knowledge and high motivation);
- *Henry, lazy expert* (high knowledge and low motivation).

Regarding Facebook's current privacy and security controls, the five-persona categorisation was found to cover the user space better than Westin's segmentation (see Dupree et al., 2016).

*Criticisms and Proposed Advancement of Westin's Approach*
The first study (2016) examined and expanded on Westin's segmentation by proposing an alternative clustering approach that reveals different categories and highlights the limitations of Westin's three-category view. The study presented a more detailed and nuanced understanding of user categorisation in relation to Westin's segmentation, emphasising the potential for improving the design of security and privacy tools. The second study (2018) acknowledged Westin's categorisation of users based on the strengths of their privacy concerns into three broad categories: the Marginally Concerned, the Privacy Fundamentalists and the Pragmatic Majority. It recognised that survey data, including Westin's work, provided the initial overview of user categories within the PAS (privacy-enhancing technologies) research domain. The paper presented a case study that aimed to develop personas representing the user space of PAS tools. It discussed the limitations of Westin's categories in predicting user behaviour and highlighted the poor performance of these categories in certain scenarios. The study indicated that a new type of categorisation was needed, which led to the development of a more refined set of personas through grounded analysis based on the dimensions of knowledge and motivation. The paper further discussed how personas generated through grounded analysis can be used to inform requirements engineering and user interface design. Finally, the study presented a gedanken experiment examining the usability of security software interfaces based on the personas' perspectives, highlighting the benefits of considering personas during design validation.

*Our Considerations*
The first study (2016) explored alternative user clustering methods, identifying distinct clusters and highlighting the limitations of Westin's segmentation.

However, the sample size was relatively small and, as also pointed by the authors, there was potential bias introduced by the use of rationales. Further, there was a lack of empirical evidence on evaluating design implications. The second study (2018) presented a validation case study, employing grounded analysis to effectively categorise the user space and create personas for requirements engineering. The analysis offered valuable insights on Westin's segmentation, highlighting its limitations and the need for alternative categorisations. The practical application of personas contributed to the understanding of user behaviours and design decisions. However, like the 2016 study, this paper had some limitations, including the limited sample of 32 subjects, mostly consisting of younger individuals. Thus, the generalisability of the results are limited. Additionally, the resource-intensive nature of the method utilised may hinder its applicability in certain contexts, and further research is needed to validate and refine the approach.

### **Schairer et al. (2019)** [40]

*Motivations*
The study aimed to develop a model of privacy disposition based on qualitative research on privacy considerations in the context of emerging health technologies.

*Research Questions*
Understanding the ways in which individuals value or do not value control over their health information, identifying motivations and deterrents related to sharing personal information that go beyond risks and benefits, examining the role of privacy philosophies as a subtype of motivation or deterrent and proposing a psychometric instrument based on the model to identify types of privacy dispositions and their applications in research, clinical practice, system design and policy.

*Methodology*
A total of 108 participants took part in the study (female = 60.2%; age range 13–82 years). The participants were recruited from various sources, including patient cohorts, community groups and online patient networks. This selection aimed to encompass a wide range of experiences, expectations and understandings of privacy in relation to emerging health technologies. The data collection involved both focus groups and individual interviews. The sessions took place over a period of several months and were conducted either in person or over the phone. Focus groups lasted for 90 minutes and were held at specific locations, while interviews had a maximum duration of 60 minutes. Focus groups were recorded using audio and video, while all interviews were audio recorded. A systematic coding process was employed to analyse the collected data. Transcripts of the focus groups and interviews were coded using thematic coding based on passages highlighting factors influencing privacy and the participants' reasons for their privacy-related decisions. A codebook was developed consisting of 27 thematic codes and eight section codes. Three independent coders applied the codes to the transcripts, with regular meetings to ensure consensus and consistency. About 19% of the transcripts were consensus coded, and inter-coder reliability was not calculated for these transcripts due to the agreement required in the consensus coding process.

*Results*

The results focussed on an analysis of 10 codes related to the disclosure of health information. These codes encompassed factors such as access control, consequences of disclosure, privacy practices, reasons for sharing (altruistic and personal), safe/unsafe information, sensitive health information, stigmatised information and 'too much information' (TMI). Based on the analysis, the researchers identified four broad categories that formed the foundation of their model of privacy disposition: 1) reasons for sharing, 2) reasons against sharing, 3) interpersonal habits and 4) institutional habits. Interpersonal habits referred to how individuals shared information with people they knew personally or encountered in person, reflecting their perceptions of privacy as a personal characteristic. Institutional habits referred to behaviours and practices related to situations where disclosed information might be recorded and used by institutions. Examples of interpersonal habits included individuals describing themselves as 'private' or 'not private' and their preferences for sharing personal health information with others. Institutional habits involved behaviours such as withholding information, lying or taking precautionary steps when sharing information with institutions. These behaviours were not always consistent with individuals' self-descriptions as private or not private, indicating that interpersonal and institutional information habits could vary independently.

The study identified various philosophies of privacy that the participants discussed when considering disclosure decisions. These philosophies included fatalism, trade-off, nothing to hide, moral right, personal responsibility and something to hide. Fatalism, trade-off and nothing to hide were often mentioned as justifications for sharing personal information, highlighting the belief that total privacy is unattainable or that the benefits outweigh the privacy concerns. Conversely, philosophies such as moral right, personal responsibility and something to hide discouraged disclosure, reflecting a higher personal value of privacy. Privacy philosophies were found to influence participants' willingness to disclose information or their selective disclosure practices. It is important to note that the participants sometimes mentioned these philosophies as beliefs held by themselves or others, providing insights on shared cultural understandings of privacy.

*Criticisms and proposed advancement to Westin's approach*

The paper offered a critique of Westin's segmentation by challenging its rigid categorisation of privacy attitudes and the limitations of focussing solely on risks and benefits. It proposed a more comprehensive conceptual model of privacy disposition, considering contextual and habitual factors, motivations and deterrents beyond risks and benefits and the inclusion of privacy philosophies. It suggested advancements in understanding and measuring privacy attitudes, advocating for a more nuanced and inclusive approach that captures the complexity of individuals' privacy-related decision-making. The paper suggested that privacy-related behaviour is both contextual and habitual, which challenges the notion of a fixed privacy attitude associated with Westin's segmentation. Thus, the study implied that individuals may exhibit different privacy behaviours and concerns depending on the specific context and their habitual patterns of information disclosure. The work expanded the understanding of motivations and deterrents related to information disclosure beyond the conventional assessment of risks and benefits. It highlighted the importance of subjective experiences, feelings, preferences and privacy philosophies on privacy-related behaviours. Overall, this critique suggested that individuals' privacy attitudes cannot be solely categorised based on concerns about risks and benefits, as proposed in Westin's segmentation. In other words, the paper argued that individuals may hold contradictory privacy philosophies and that these philosophies may vary among individuals, challenging the rigid categorisation of privacy attitudes in Westin's segmentation.

*Our Considerations*
The work adopted a comprehensive qualitative research approach, involving a diverse range of participants and employing rigorous coding and analysis techniques. It addressed the limitations of Westin's segmentation, proposing a more nuanced conceptual model of privacy disposition that incorporates contextual and habitual factors, motivations and deterrents beyond risks and benefits and includes privacy philosophies. This advancement in understanding privacy attitudes has implications for research, clinical practice, system design and policy. As acknowledged by the authors themselves in the Limitations section, the study included a non-representative sample, which may have affected the generalisability of the findings. Additionally, while the qualitative analysis offered rich insights, future research is needed to quantitatively validate and operationalise the proposed model. The paper acknowledged the ethical implications of privacy and information disclosure in the context of emerging health technologies. It highlighted the importance of informed consent processes, user-centered approaches and the development of tailored decision aids to address privacy concerns. By exploring individuals' privacy dispositions and considering their values and expectations, the paper aimed to contribute to more ethical practices in research, health care and policy.

**Toresson et al. (2020)** [45]

Motivations
Creating an educative self-assessment app named PISA to increase the awareness of app-related privacy risks.

*Research Questions*
The purpose of this study was to evaluate the effectiveness of privacy impact self-assessment (PISA) apps.

*Methodology*
1) Creation of static app identification data from the KAUDROID database, which provides Android app permission statistics.
2) Development of personas with specific privacy vulnerabilities.
3) Mapping of identification/de-anonymisation threats to each persona's vulnerabilities.
4) Definition of the privacy impact of each realised threat on each persona.

The privacy threats were modelled as identification risks based on data shared through apps. The assumption was that individuals who can be partially identified through certain attributes are exposed to privacy risks. The model utilised the KAUDROID data to create a record for each app, describing the identity attributes accessed by that app. Privacy impact and data protection impact analyses were conducted to determine the privacy impact. The PRIAM method categorises privacy harms into five categories: physical harms, economic/financial harms, mental/psychological harms, harms to dignity/reputation and societal/architectural harms. ENISA offers a similar conceptualisation with impact levels described as low, medium, high, and very high. The impact definitions in these two frameworks were used to define the impact levels in the research. Additionally, a walk-through of the PISA user interface is provided, illustrating the interactions and steps involved in using the app. The walk-through demonstrates the greeting screen, app selection, persona selection, persona details, privacy impact information, and the final mitigation step. The methodology incorporates data analysis, persona creation, threat mapping, impact definition, and the development of an interactive user interface to support the research objectives of the PISA app.

*Results*
Five personas were created/described (based on Dupree et al.'s classification), considering specific privacy vulnerabilities (or threats) in life contexts. Using the PISA app, users could randomly select an installed app from a database (KAUDROID) of app permission statistics of permissions used to access data on phones and then choose one of the five vulnerable personas, select partial consent and provide a mitigation action aimed at reducing privacy vulnerabilities. Thus, the intention of the PISA app was to increase users' awareness of data sharing and risks while installing apps, using concrete examples of vulnerable personas:

- female, e-sport celebrity, using a pseudonym (stalking, sabotage, sponsor loss);
- male, well-off elderly citizen with early dementia (exploitation, fraud, social exclusion);
- male, mid-life professional career, undergoing, cancer treatment (career

damage, relationship distress, abusive phone sellers);

- male, married, regional politician with a predilection for extramarital affairs (public and private trust engendered, divorce, economic loss);
- teenage, female homosexual in intolerant social environment (discrimination, exclusion, risky contact proposal).

The study successfully achieved its goal of creating a swipe-friendly user interface and received positive informal feedback on the app. However, some practical issues were identified, including the limited number of personas and the restriction of interactions to the apps contained in the KAUDROID database.

*Criticisms and Proposed Advancement of Westin's Approach*
The paper implicitly related to Westin's segmentation by incorporating personas with specific vulnerabilities, which aligns with the idea of categorising individuals based on their privacy concerns and behaviours. By utilising personas to represent different segments of smartphone users with varying privacy vulnerabilities, the paper acknowledged the variability in privacy attitudes and recognised the need for personalised approaches to privacy management. The focus on engaging users in reflecting on their data-sharing behaviours and privacy risks aligned with the goals of understanding and addressing individual differences in privacy concerns, which are central to Westin's segmentation framework. While the paper did not explicitly discuss or critique Westin's segmentation, its utilisation of personas underscores the importance of recognising and accommodating individuals' diverse privacy needs.

*Our Considerations*
The paper presented an app that promotes reflection on data-sharing and privacy-related risks among smartphone users. The incorporation of personas adds a personalised dimension to the app, allowing users to relate to specific vulnerabilities. However, the paper also had some weaknesses. The procedure that was used to create the personas was not fully elaborated. In addition, as acknowledged by the authors themselves, the limited number of personas and the restriction to the apps included in the KAUDROID database hinder the app's coverage and its ability to address a wider population. Also, the static inclusion of app statistics might not be adaptable to evolving app behaviours, potentially limiting the app's accuracy over time.

**Di Ruscio et al. (2022)** [16]

Motivations
The authors analysed the possibility of building profiles from answers to general questions and predicting privacy preferences using those profiles through the use of a recommender system.

*Research Questions*
The paper aimed to identify relevant sets of general privacy questions to classify users based on their moral privacy preferences. The work examined the alignment between users' self-assessment of privacy attitudes and their actual behaviours in practice. The authors also developed a recommender system, PisaRec, to offer privacy settings that reflect user preferences.

*Methodology*
This paper used an existing data set on fitness app usage. User privacy preferences were utilised for the evaluation of the system, consisting of domain-specific, app-related and generic questions, and an evaluation of the proposed approach was presented based on several metrics. Generalisable questions were extracted through a qualitative analysis from the original data set, and the results were subjected to multiple comparisons. A compactness metric was used to measure the relevance of users within a cluster, with lower values indicating better clustering solutions. A silhouette metric was used to assess the similarity of a user to others in the same cluster, with higher scores indicating better clustering. Precision and recall were used to evaluate the classification of recommended settings compared to ground-truth data. The false positive rate (FPR) was used as a measure of the ratio of falsely classified items. Additionally, the performance was analysed using the receiver operating characteristic (ROC) curve and the area under the curve (AUC), with an ROC close to the upper-left corner indicating better prediction performance. The methodology involved user privacy profiling, clustering, classification and recommendation of privacy settings. It leveraged both supervised and unsupervised techniques as well as a collaborative-filtering approach to provide personalised automated privacy assistance to users.

*Results*
In agreement with the Privacy paradox, the results showed that users' self-assessments of their privacy category did not align with their actual privacy category. Users with similar privacy settings perceived themselves as belonging to different groups, leading to the low prediction accuracy of the neural network classifier. The evaluation of different sets of questions revealed that a combination of generic questions and generalisable ones provided the best clustering solution for assessing privacy concerns. The performance of PisaRec, the privacy settings assistant, was validated, as the recommender system effectively recommended relevant settings to users, even with a limited amount of training data. The prediction performance of PisaRec improved as more data were made available for training. The results indicate the importance of considering user privacy preferences beyond self-assessment and highlighted the efficacy of the proposed methodology in categorising users, assessing privacy concerns and recommending personalised privacy settings. The findings confirm the usefulness for automated privacy assistance to mitigate the inconsistencies in users' self-perceptions and provide tailored privacy solutions.

*Criticisms and Proposed Advancement of Westin's Approach*
The paper relates to Westin's segmentation of user profiles introducing a new categorisation framework. It is composed of new categories (Inattentive, Involved/Attentive and Solicitous), allowing for a classification of users based on their general privacy preferences and attitudes. It also presented an automated approach to creating user privacy profiles, leveraging unsupervised clustering and a graph-based representation of users and their privacy settings. These advancements enable a more comprehensive and accurate classification of user privacy profiles and the automated profiling with the personalised privacy setting recommendations were shown to promote better privacy management. The advancements proposed in the paper contribute to a more sophisticated understanding of user privacy profiles and improve privacy management practices.

*Our Considerations*
The automated approach to creating user profiles based on unsupervised clustering and a feed-forward neural network is a strength of this work, improving classification accuracy. The introduction of PisaRec, a privacy settings assistant powered by a recommender system, also added value by providing personalised privacy setting recommendations. However, the study had some weaknesses, including the limited data set used for evaluation and the suboptimal performance of the neural network classifier in predicting users' self-assessed privacy categories. Additionally, a more thorough validation of the proposed methodology is needed using larger and more diverse data sets.

## 5. Discussion

In the following, the three research questions are discussed in relation to the literature reviewed and the criticisms that emerged.

### 5.1. Research Questions

**RQ1**: Identify the **study contexts** that propose privacy categorisations.

It is possible to determine the study contexts based on the motivations and research questions presented in the reviewed papers. We can divide the approaches into two families, which are also temporally distinct. The first one involves the investigation of privacy profiles to better characterise online users in order to provide improved/customised content or services. This first family regards online users as *consumers*. The second one utilises privacy profiles as a piece of information that can better express the privacy characteristics of an online user and can be used to empower the user's activity in the digital world. This second family considers the user a citizen of the digital world whose rights need to be protected. These two approaches are temporally distributed in the time period of the considered papers, from the early 1990s to the early 2020s, which can be further divided into three periods. In the first period, 1991–2005, papers primarily focused on users' privacy from a consumer perspective, with marketing and e-commerce being the dominant application domains. In 2005, Consolvo's work marked a shift toward understanding user privacy preferences in relation to the the development of location-based services. Studies in this period also acknowledged the user as both a consumer and a social actor in the digital world, reflecting the increased integration of digital experiences in everyday life.

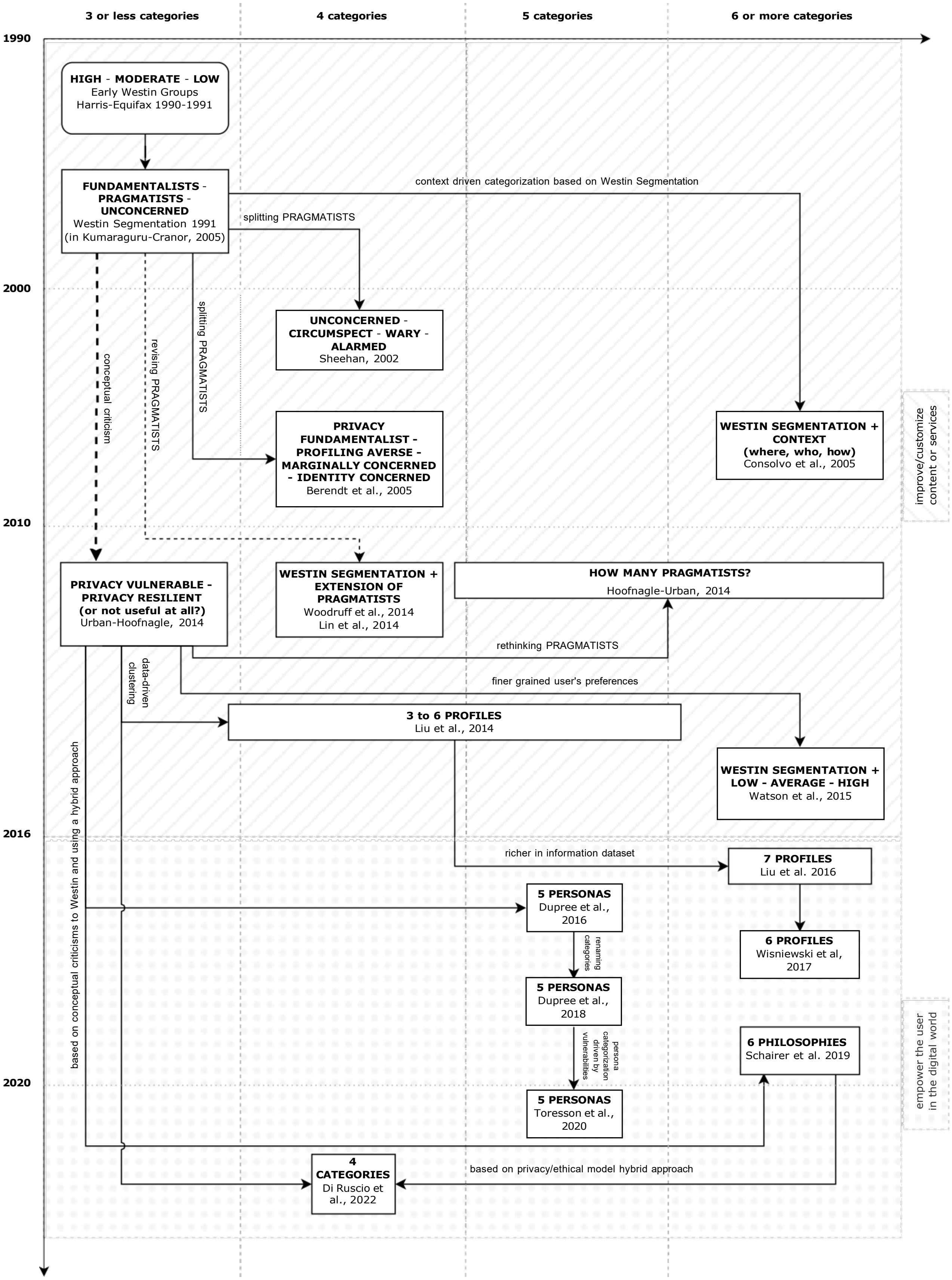

Figure 1: Evolution map of privacy categorisation

The second period, spanning until 2016, saw significant criticism of the economically driven definition of online user privacy [22, 47, 61]. During this time, new approaches emerged that sought to understand and support users' privacy settings and management based on their own attitudes [51, 26, 29].

In 2016, a paradigm shift occurred, with a focus on considering the user as a digital citizen. This shift involved characterising privacy attitudes using personas [17, 18], philosophies [40] and profiles that empower and support user interactions in the digital realm [59, 45, 16].

Figure 2 and Table 3 present the **nine main contexts/domains** in which the privacy categorisation was used or referenced: *General, Economy, Mobile Apps, Health, Social networks, Computing, Location sharing, E-commerce, Internet*. With general we mean that the investigation although conducted in a specific context, achieved general, domain-independent considerations on privacy.

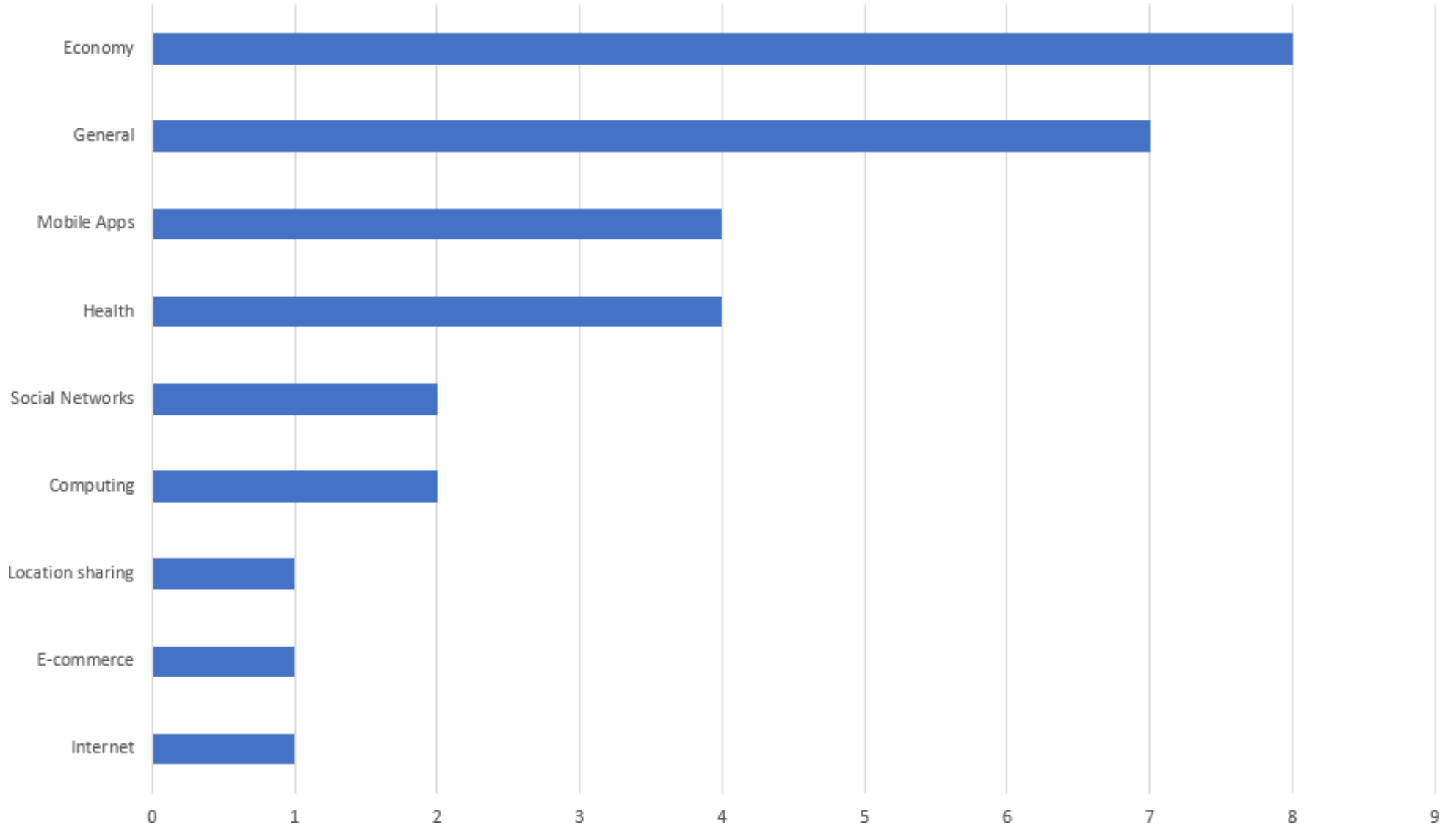

Figure 2: Domains/contexts of the studies

**RQ2**: Understand the **methodologies** and **approaches** of privacy categorisations.

We observed three different approaches to privacy categorisation. We refer to them as follows:

- **Model-Driven**
- **Data-Driven**
- **Hybrid**

The primary focus of a **model-driven** approach is the creation and use of domain models, which are conceptual representations of the subjects relevant to a particular issue. Instead of focussing on the actual data, it places attention on abstract representations of the knowledge and activities that regulate a specific application domain.

An approach that is **data-driven** indicates that the choices are based on the analysis and interpretation of data. A data-driven method helps to avoid the introduction of bias into the study due to the researcher's own experiences or existing theories.

In addition to these two approaches, we repeatedly observed the use of a **hybrid** approach, which involves the application of both data-driven and model-driven aspects of research — often in that order. In the hybrid approach, further interpretative proposals are guided by models emerging from the data. This hybrid approach can also incorporate a **grounded analysis** based on grounded theory [33], in which inductive reasoning is used in the process of developing the analysis.

Table 3 provides a synthetic view of the different approaches.

Based on the approaches described above, different methodologies are applied to obtain the categorisations:

- **Segmentation** is the process of partitioning a data set into meaningful regions or extracting relevant features from it [60]; this process is mainly model driven and is used to create segments.
- **Clustering** is the process of mathematically grouping similar objects into different groups [31]; it is mainly data driven, although it can also include modelling, and it is used to create clusters.
- **Profile/Profiling** involves mostly hybrid approaches. Specifically, it relies on correlated data created using different methodologies to identify and represent a subject (individual or group) [21, 39]. The correlated data aggregation involves different sources, and individuals are usually not aware of this process. The group profiling process can be distributive (e.g. the same characteristics apply to both the group and all its members) or non-distributive (the attributes of the group do not apply to all the members, and the association is statistical rather than determinate) [39].

- **Persona**, researchers usually utilise a hybrid approach (personification) that involves inducing and attributing further parameters to existing segments or clusters. This means that personification makes use of data and model-driven methods in conjunction with grounded analysis [17]. Indeed, a digital persona is "a model of an individual's public personality based on data and maintained by transactions, and intended for use as a proxy for the individual" [11, 39].

Regarding the instruments, most of the studies used questionnaires to collect data about privacy behaviours or privacy permissions. Data set analysis, also based on previous studies, and interviews utilising open questions were also adopted, whereas other instruments, such as focus groups, self-assessments of apps, simulations or literature reviews, were rarely used. Instead, data collection from existing databases was more commonly used. (Fig.3).

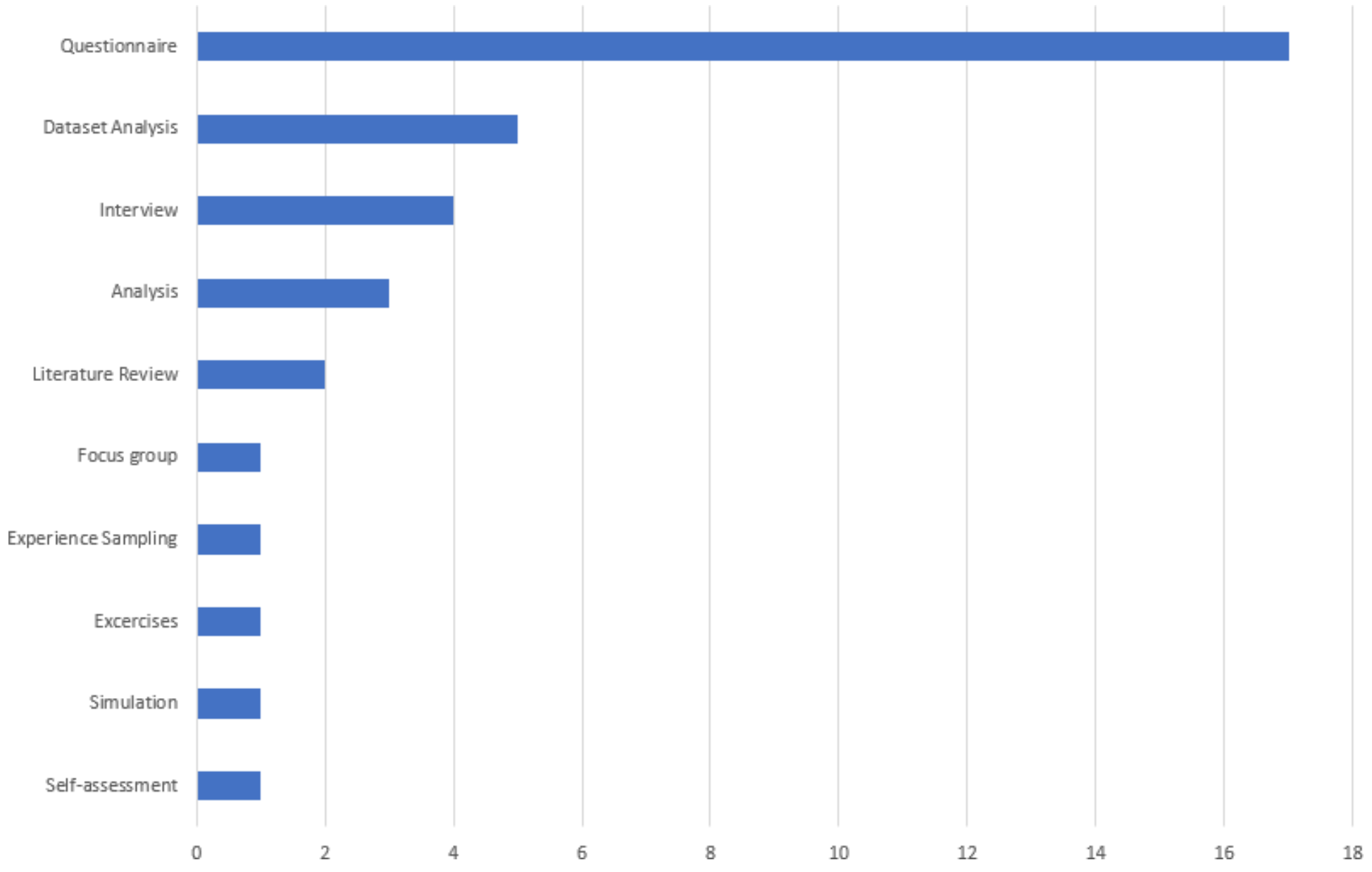

Figure 3: Instruments used

The categories personas/philosophies created using hybrid approaches (data- and model-driven approaches, e.g. clustering + profiling, and grounded analysis, e.g. inductive reasoning) produce a more detailed and specific categorisation than data clusters, modelled segments or parameter-driven profiles. Indeed, the hybrid approach may balance the limits of the used approaches. Notably, only Liu et al. [28] used a purely data-driven approach. Such an approach may suffer from the limitations and biases of the analysed sample, being conditioned by its size [13, 29], homogeneity (e.g. subjects with medical conditions [40]) and representativeness (e.g. of a certain country [28]). In fact, most of the reviewed studies were not balanced in terms of gender [52, 40, 59], and the factors of age, education level [17, 59] and technology proficiency were not appropriately used as controlling factors for privacy behaviours. Recently, the introduction of modern data analysis methodologies has resulted in the consolidation of some categorisations at the expense of others [13, 61]. For example, Dupree [17] and Schairer [40] used both hybrid and interdisciplinary approaches, including the human science perspective, to fully analyse and manage users' digital privacy.

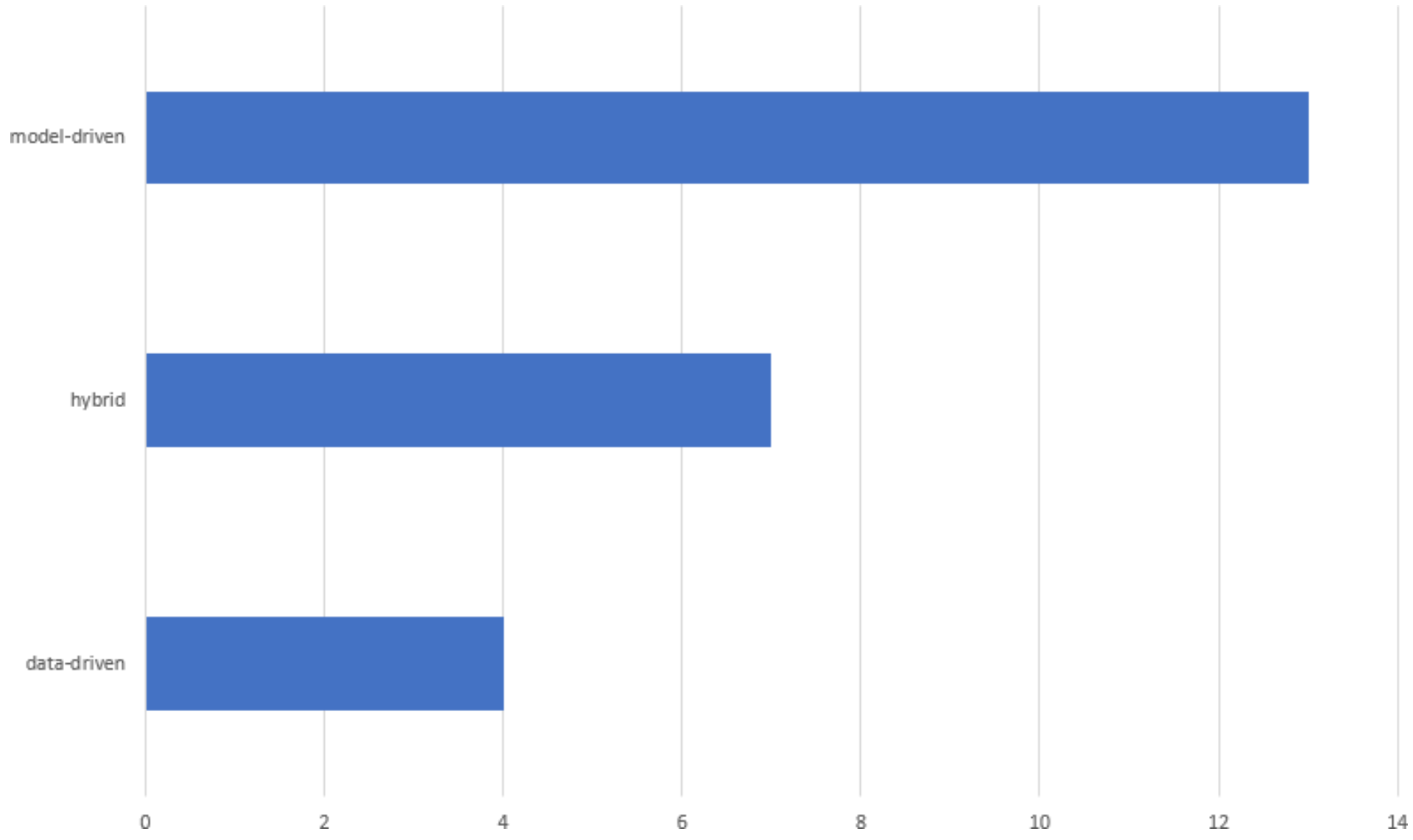

Figure 4: Approaches used

**RQ3**: Mapping the **evolution** of the privacy categorisations and **definitions** of the categories.

The graphical representation in Fig.1 maps a temporal sequence of the evolution of privacy categorisation research from 1991 to 2022. The temporal dimension is represented on the y-axis, enabling the tracking of the evolution of privacy categorisation over time. The x-axis of the graph represents the varying number of categories included in each study, with a range of three or fewer to six or more. This reflects the progressive increase in the detail of privacy categorisation over the course of time. Starting with Westin's original idea of measuring privacy awareness (reported in the rounded box), the diagramme comprises sixteen distinct squared boxes, each denoting a particular study and research endeavour pertaining to the classification of privacy. Additionally, the graph incorporates a background fill, which serves to highlight a change in the predominant research interests of the community over the course of time. The hatched portion of the graph spanning from 1991 to 2016 denotes research endeavours motivated by the need to improve or customise content or services. The period denoted by the dotted fill, spanning 2016 to 2022, highlights a transition in research motivated by the objective to empower users in the digital world. The boxes are linked by lines of diverse typologies, serving as representations of the heterogeneous character of the conceptual associations among the studies: filled lines suggest a conceptual linkage between two studies, indicating that the subsequent research was either influenced by, built on or shared the same conceptual approach as the preceding one; the use of dotted lines serves as a representation of a conceptual deviation, indicating that the ensuing investigation introduced modifications to the initial methodology or concept; and the presence of bold dotted lines serves as a clear indication of a significant conceptual departure and critical evaluation of the preceding study, thus denoting a noteworthy alteration in the research perspective. The starting point of the journey can be traced back to the year 1991, when Westin introduced its Privacy Segmentation Index (reported by Kumaraguru-Cranor, 2005), which is symbolised by the initial squared box. This research constitutes the fundamental groundwork in the domain of privacy classification. Westin's research has led to the emergence of several divergent studies. Three authors (Sheehan, 2002; Consolvo et al., 2005; Berendt et al., 2005) have expanded on or modified Westin's methodology and thus are connected by solid lines, respectively. The Woodruff et al. (2014) and Lin et al. (2014) box is linked to the Westin Segmentation box using a dotted line, indicating the conceptual revision of the *pragmatists* group. Urban-Hoofnagle's (2014) findings, connected using a bold dotted line, represented a noteworthy departure from Westin's segmentation and posed a critical challenge to it. The network derived by the Urban-Hoofnagle (2014) box reveals all the subsequent series of works in the following years, all linked by filled lines. The nearest connected network nodes comprises Hoofnagle-Urban (2014), Liu et al. (2014) and Watson et al. (2015). Notably, Liu et al. (2014) employed a purely data-driven approach using clustering methodologies, in the same way as Liu et al. (2016), to which it is connected. In a similar way, Wisniewski et al. (2017) is connected to it since the management profiles of privacy behaviour are identified using clustering. Also taking into account the Urban-Hoofnagle departure from Westin segmentation, the work of Dupree et al. (2016) proposed *personas* adopting a hybrid approach, leading to Dupree et al. (2018) and Toresson et al. (2020), which categorised personas using privacy vulnerabilities. Also connected to the Urban-Hoofnagle (2014) box is Schairer et al. (2019), who applied qualitative research to elucidate *philosophies* of privacy. The last in chronological order is Di Ruscio et al. (2022), which, based on Urban-Hoofnagle's (2014) critiques and Schairer et al.'s (2019) qualitative methodology, refines the categorisation and proposes four privacy categories. In conclusion, the evolution map graph shown in Fig.1 serves as a chronological and visual representation of the categorisation of privacy development. Specifically, it illustrates the evolution of concepts, methodologies and approaches as well as the shifting research focus over time.

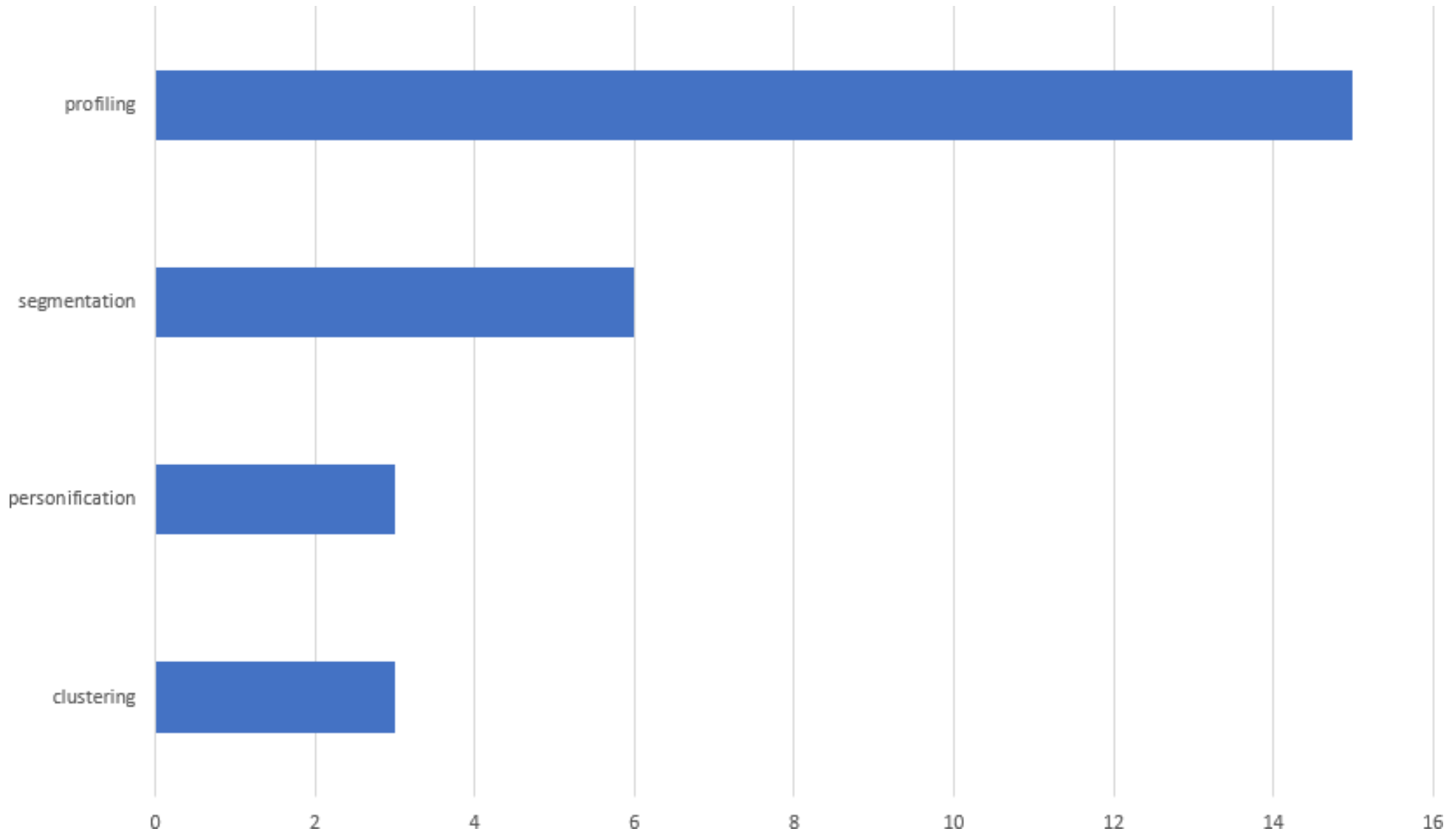

Figure 5: Categorisations used

## 6. Conclusions

In the realm of digital privacy, the concept of privacy categorisation has gained increasing prominence in recent years due to the widespread integration of information technology in our daily lives. This systematic review investigates the issue of privacy categorisation, focusing on five keywords: profile, profiling, segmentation, clustering and personae. These keywords were identified through a comprehensive analysis of previous studies and references, covering the fundamental concepts and methodologies employed in the creation of privacy categories. The goal is to provide a thorough investigation into the landscape of privacy categorisation.

Analysing the temporal evolution of the definitions related to the categorisations, the terms initially chosen reflected more vagueness (e.g. Westin [53]), conveying a blurred vision of privacy. In contrast, earlier works published from 2000 to 2016 predominantly focused on a technology-oriented perspective of privacy (e.g, Berendt [8], Wisniewski [59]). However, recent works [28, 17, 18, 40, 45, 59, 16] demonstrate a shift toward the use of privacy preferences to support users in their interactions with digital systems. These works also explore broader concepts, such as philosophies of privacy versus app privacy settings, thus touching on the ethical dimension of privacy. Examining the world cloud in Fig. 6, it is evident that terms like *privacy*, *user/users*, *information*, *participants* and *data* are prominent. However, there is also a clear presence of words like *app/apps*, *settings*, *preferences*, *permissions* and *location*, *consumer*. Terms like *attitudes*, *behaviours* and *ethics* appear less frequently, with the latter not surpassing the threshold.

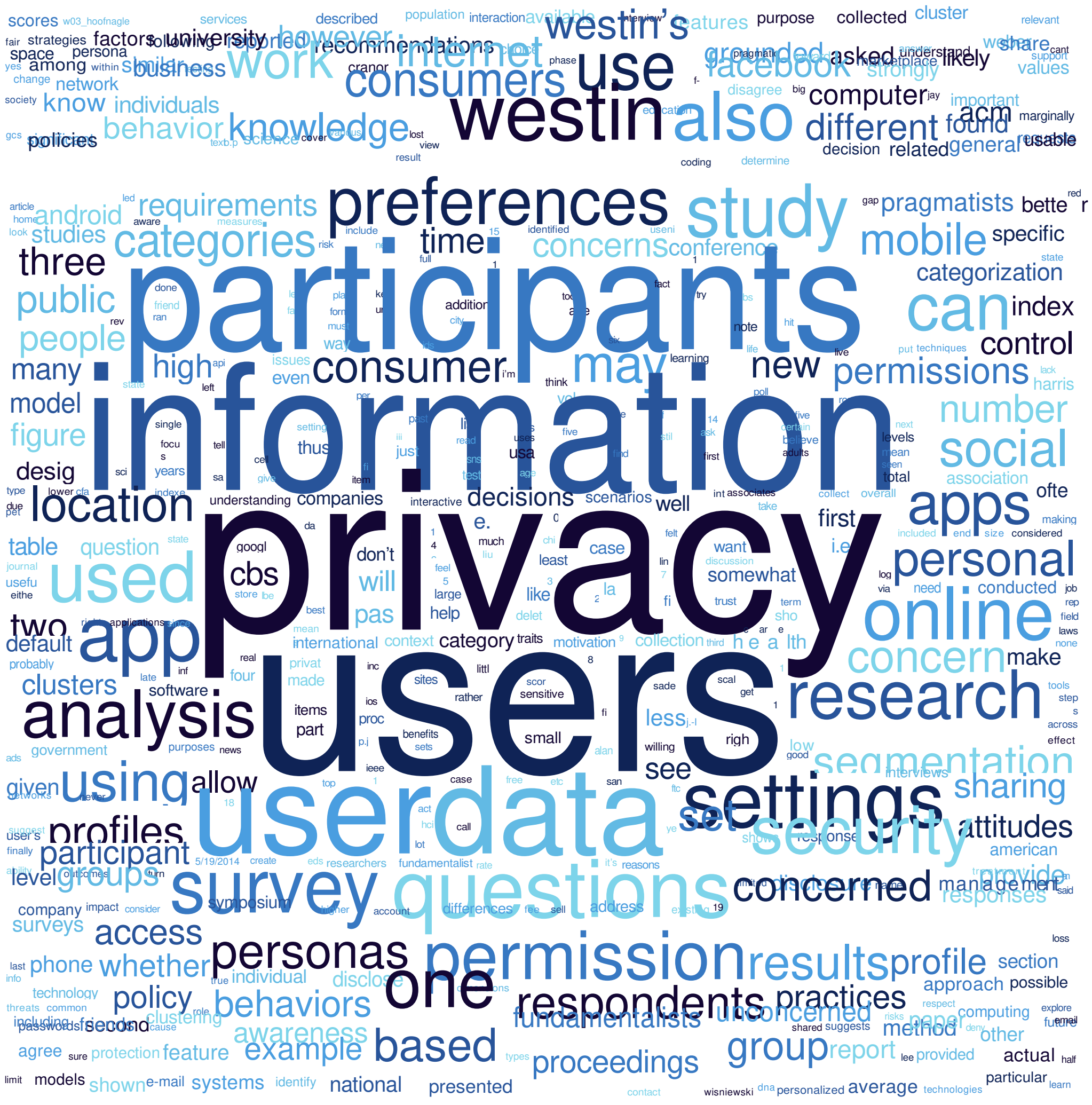

Figure 6: Overview of the keywords found in the papers analysing the whole text, shown with size relative to frequency

The survey reveals that categorisations has evolved in terms of both their meanings and numbers. The complexity managing user privacy in an increasingly digitalised world has made it impractical to establish a fixed number of categories. This raises the question of whether privacy categorisation as a research endeavour remains meaningful and has a future.

As mentioned above, the evolution of digital technology and its diffusion in everyday life [19] seems to demand a more holistic conception of privacy. Although recent regulations (e.g. GDPR [38]) require that any product or service must obtain consent from users regarding the way in which their data will be managed (e.g. by websites and their third parties), the power asymmetry between users and digital systems leaves the former unprotected in terms of privacy and security. A focus is needed on the ethics of digital technology beyond the concept of 'online privacy', orienting toward a vision of privacy as a fundamental right of human beings in order to preserve the human agency, autonomy and dignity. The latter implies «the recognition of the inherent human state of being worthy of respect», which «must not be violated by 'autonomous' technologies» [37, 4]. Accordingly, approaches that empower humans in their interactions with autonomous system by exploiting their ethical preferences, such as [6], as well as approaches that leverage human values in software engineering, such as [41], have appeared recently.

These attempts may help to provide a better understanding of the notion and practical purpose of privacy categorisations. Indeed, although it is important to preserve the peculiarity of specific domains, categorisation could adopt a more domain-independent approach to better reflect privacy preferences in terms of users' ethical/moral dispositions. In other words, based on the idea that privacy relies on abstract principles and reflects personal ethics, categorisation and any recommendation means based on it could be driven by individual dispositions (e.g. personality traits, attitudes), world views (e.g. normativism and humanism), ethical considerations (e.g. ethical ideologies, such as idealism and relativism) and so on [4] rather than by contextual factors or specific attitudes and practices in a given domain [16]. This suggests that privacy preferences could be defined at two levels: the domain-specific level, reflecting the norms, rules and values conveyed by the domain (e.g. the SNS, which involves, for example, sharing locations and photos or apps that gather health-related data), and the domain-general level, reflecting behaviours and preferences at a more abstract level, applying to a variety of situations or contexts. This also suggests that in order to manage this increasing complexity in categorisation and develop a faithful characterisation of user privacy and ethical behaviour, a continuous learning process is needed that includes monitoring user behaviour and providing control feedback to users when offering recommendations based on user profiles and making automatic decisions on behalf of the user. Thus, categorisation may be helpful in positioning the users [28] and interpreting the feedback while learning.

In summary, the investigation of digital privacy behaviours and categorisations has several theoretical and practical implications:

- to relate the new approaches and emerging digital categories to Westin's seminal work.

- to understand the complexity of digital privacy behaviour, which is subject to continuous changes and adaptations due to new regulations, opportunities, digital skills and societal awareness.

- to adapt the digital privacy protocols and new technologies to categories in order to satisfy the new challenges and demands of society. This implies better design of digital technologies based on users' digital privacy preferences.

- to encourage digital privacy education, improving users' knowledge and proficiency as well as making easier to engage in a specific behaviour.

In conclusion, privacy categorisation appears to have great potential to help manage the digital privacy issues that users encounter when interacting with increasingly autonomous systems. To better explore the opportunities of privacy categorisation and the recommendation systems that may arise from it, future works should examine privacy categories on a more general level using a variety of multi-disciplinary approaches and perspectives. Further, research is needed on more descriptive categories, such as *personas* and *philosophies*. Such research could enhance the correspondence between users' general privacy beliefs and their behaviours when expressing privacy preferences.

Table 3:

| STUDY | DOMAIN | INSTRUMENTS | APPROACH | CLASSIFICATION ELEMENTS | SUBJECTS |
|---|---|---|---|---|---|
| Westin (1990) | Economy | Analysis + Questionnaire (4 Questions) | Model-Driven | (3) General Privacy Concern Index: High, Moderate, Low | 2254 |
| Westin (1991) | Economy | Questionnaire (4 Questions) | Model Driven | (3) Consumer Privacy Concern Index: High (Fundamentalists), Moderate (Pragmatic), Low (Unconcerned) | 1255 |
| Westin (1993) | Health | Questionnaire (4 Questions) | Model Driven | (3) Medical Privacy Concern Index: High, Medium, Low | 1000 |
| Westin (1993) | Health Computing | Questionnaire (3 Questions) | Model driven | (3) Computer Fear Index: High Computer Fear, Medium Computer Fear, Low Computer Fear | 1000 |
| Westin (1994) | Economy | Questionnaire (4 Questions) | Model Driven | (4) Distrust Index: High Distrust, Medium Distrust, Low Distrust, No Distrust | 1005 |
| Westin (1996) | Economy | Questionnaire (4 Questions) | Model Driven | (3) Privacy Concern Index: Privacy Fundamentalists, Privacy Pragmatists, Privacy Unconcerned | 1005 |

Continued on next page

Table 3: (Continued)

| | | | | | |
|---|---|---|---|---|---|
| Westin (2001) | Economy | Questionnaire (3 Questions) | Model Driven | (3) Privacy Segmentation Index: Privacy Fundamentalists, Privacy Pragmatists, Privacy Unconcerned | 1529 |
| Kumaraguru, Cranor (2005) | Economy | Literature Survey | Model Driven | (3) Privacy Segmentation, Index: Privacy Fundamentalists, Privacy Pragmatists, Privacy Unconcerned | 1529 |
| Sheehan (2002) | Internet Usage | Questionnaire (15+8 Questions) | Model Driven | (3) Privacy Segmentation Index + Typology of Internet Users: Unconcerned, Circumspect, Wary, Alarmed | 889 |
| Berendt, Gunther, Spiekermann (2005) | E-Commerce | Questionnaire (56 Questions) + Simulation | Model Driven | (4) Personal Consumer Information Cost Index: Privacy Fundamentalists, Profiling Averse, Marginally Concerned, Identity Concerned | 171 |
| Consolvo, Smith, et al. (2005) | Location Sharing | 3 Phases: 1) Questionnaire and Exercises, 2) Experience Sampling Method, 3) Interview | Model Driven | (3) Privacy Segmentation Index: Privacy Fundamentalists, Privacy Pragmatists, Privacy Unconcerned + Who was Requesting, Why They Wanted, How User Feels | 16 |
| Urban, Hoofnagle (2014) | Economy (General) | Surveys and Analysis | Hybrid | (2) Privacy Vulnerable, Privacy Resilient (or not useful at all?) | 2203 |
| Hoofnagle, Urban (2014) | Economy (General) | Surveys and Analysis | Hybrid | (3) Privacy Segmentation Index Critics: Privacy Fundamentalists, Privacy Pragmatists (to be revised), Privacy Unconcerned | 2203 |
| Woodruff, Pihur, et al. (2014) | Health Privacy (General) | Questionnaire | Model Driven | (4) Privacy Segmentation Index Critics: Fundamentalists, Pragmatists, Unconcerned + 'Fundamentalists Pragmatists' | 884 |
| Liu, Lin, Sadeh (2014) | Mobile Apps | LBE Privacy Guard Dataset Analysis | Data Driven | from 3 to 6 profiles (dataset dependent) | 4.8M |
| Lin, Liu, et al. (2014) | Mobile Apps | Google Play API Data Analysis | Data Driven | 4 profiles: Conservatives, Unconcerned, Fence-Sitters , Advanced Users | 725 |
| Watson, Lipford, Besmer (2015) | Social Network Service: Facebook | Survey for the Usage and General Privacy Attitudes: Westin's Questions; Buchanan Index; Facebook Intensity Index | Model Driven | (3) Westin's Profiles (pragmatist, fundamentalist, unconcerned) + Low, Average, High | 184 |
| Liu, Andersen, et al. (2016) | Mobile Apps | Enhanced Android Permission Manager Dataset Analysis | Data Driven | 7 profiles (dataset dependent) | 72 |

Continued on next page

Table 3: (Continued)

| | | | | | |
|---|---|---|---|---|---|
| Dupree, DeVries, et al. (2016) | General | Survey and Open-ended Interviews | Hybrid | 5 Personas: Marginally Concerned, Fundamentalists, Amateurs, Technicians, Lazy Experts | 212 |
| Wisniewski, Knijnenburg, Lipford (2017) | Social Network Service: Facebook | Survey for Privacy Behaviours (management) and Feature Awareness (proficiency) | Data Driven | 6 Management Profiles for Privacy Behaviour: Privacy Maximizers, Self-Censors, Time Savers/Consumers, Privacy Balancers, Selective Sharers, Privacy Minimalists - Six Proficiency Profiles for Feature Awareness: Novices, Near-Novices, Mostly Novices, Some Expertise, Near Expertise, Experts | 308 |
| Dupree, Lank, Berry (2018) | Computer Based System: General | Open-ended Interviews | Hybrid | 5 Personas: Marginally Aware, Fundamentalists, Struggling Amateurs, Technicians, Lazy Experts | 32 |
| Schairer, Cheung, et al. (2019) | Health Privacy (General) | Focus Group, Interview,Questionnaire | Hybrid | (6) Philosophies of Privacy: Fatalism, Moral Right, Nothing to Hide, Something to Hide, Personal Responsibility, Trade-off | 108 |
| Toresson, Shaker, et al. (2020) | KAUDroid dataset | PISA (privacy impact self-assessment) App | Hybrid | 5 Personas (descriptive) | n/a |
| Di Ruscio, Inverardi, et al. (2022) | General | Cross-Domain Dataset Analysis | Hybrid | (3-4) Inattentive, Involved/Attentive, Solicitous | 295 |

Table 3: Reviewed articles with research details